\documentclass[11pt]{article}
\usepackage{fullpage}
\usepackage{amsmath}
\usepackage{amssymb}
\usepackage{amsfonts}
\usepackage{epsfig}
\usepackage[all,poly]{xy}
\usepackage{colortbl}

\usepackage{geometry}
\usepackage{tikz}
\usepackage{mathrsfs}

\newif\ifnotesw\noteswtrue% T to show box & marginal notes; F suppresses.
   {\ifnotesw\marginpar[\hfill\(\top\)]{\(\top\)}\fi}%
   {\ifnotesw\marginpar[\hfill\(\bot\)]{\(\bot\)}\fi}

\newcommand{\mnote}[1]%
    {\ifnotesw\marginpar%
        [{\scriptsize\begin{minipage}[t]{\marginparwidth}
        \raggedleft#1%
                        \end{minipage}}]%
        {\scriptsize\begin{minipage}[t]{\marginparwidth}
        \raggedright#1%
                        \end{minipage}}%
    \fi}
\newcommand{\ignore}[1]{}

\newcommand{\etal}{{\it et al.~}}

\newsavebox{\given}
\savebox{\given}[1em]{\rule[-1.5ex]{.2mm}{4ex}}

\newcommand{\bnum}{\begin{equation}}
\newcommand{\enum}{\end{equation}}

\newtheorem{theorem}{Theorem}
\newtheorem{corollary}[theorem]{Corollary}
\newtheorem{lemma}[theorem]{Lemma}

\newtheorem{fact}[theorem]{Fact}

\newtheorem{conjecture}{Conjecture}

\newtheorem{definition}{Definition}

\newcommand{\blackslug}{\rule{7pt}{7pt}}

\newcommand{\iverson}[1]{\lbrack\!\lbrack #1 \rbrack\!\rbrack}
\newcommand{\qed}{\hfill{\setlength{\fboxsep}{0pt}
\framebox[7pt]{\rule{0pt}{7pt}}}}

\renewcommand{\notin}{\ifmmode \not\in \else $\not\in$ \fi}
\newlength{\thislabel}
\newcommand{\labsize}[1]{\settowidth{\thislabel}{#1}}

\newcommand{\prf}{\par\noindent{\sl Proof } \hspace{.01 in}}

\newcommand{\lip}{\langle}
\newcommand{\rip}{\rangle}
\newcommand{\NN}{\mathbb{N}}
\newcommand{\CC}{\mathbb{C}}
\newcommand{\ZZ}{\mathbb{Z}}
\newcommand{\RR}{\mathbb{R}}
\newcommand{\SSS}{\mathbb{S}}

\newcommand{\GG}{\mathcal{G}}
\newcommand{\OO}{\mathcal{O}}

\newcommand{\bra}[1]{\lip #1 |}
\newcommand{\ket}[1]{| #1 \rip}

\newcommand{\ketbra}[2]{| #1 \rip\lip #2 |}

\newcommand{\cart}{\mbox{ $\Box$ }}
\newcommand{\ecart}{\mbox{\scriptsize $\square$}}

\newcommand{\AS}{\mathscr{A}}
\newcommand{\BS}{\mathscr{B}}

\newcommand{\HS}{\mathcal{H}}
\newcommand{\PS}{\mathscr{P}}

\newcommand{\ZS}{\mathcal{Z}}

\DeclareMathOperator{\Aut}{Aut}
\DeclareMathOperator{\diag}{diag}
\DeclareMathOperator{\Circ}{Circ}
\DeclareMathOperator{\Stab}{Stab}

\newcommand{\FF}[2]{{#1}^{\odot {#2}}}

\title{
Perfect state transfer on quotient graphs
}
\author{
Rachel Bachman\\{\em Clarkson University}
\and
Eric Fredette\\{\em Clarkson University}
\and 
Jessica Fuller\\{\em Seton Hall University}
\and 
Michael Landry\\{\em UC Berkeley}
\and
Michael Opperman\\{\em Clarkson University}
\and
Christino Tamon\footnote{Contact author: tino@clarkson.edu}\\{\em Clarkson University}
\and
Andrew Tollefson\\{\em University of Nevada at Reno}
}
\date{\today}
\begin{document}
\maketitle
\bibliographystyle{plain}

\begin{abstract}
We prove new results on perfect state transfer of quantum walks on quotient graphs.
Since a graph $G$ has perfect state transfer if and only if its quotient $G/\pi$, under any
equitable partition $\pi$, has perfect state transfer, 
we exhibit graphs with perfect state transfer between two vertices but which lack automorphism swapping them.
This answers a question of Godsil ({\em Discrete Mathematics} {\bf 312}(1):129-147, 2011).
We also show that the Cartesian product of quotient graphs $\Box_{k} G_{k}/\pi_{k}$
is isomorphic to the quotient graph $\Box_{k} G_{k}/\pi$, for some equitable partition $\pi$.
This provides an algebraic description of a construction due to Feder
({\em Physical Review Letters} {\bf 97}, 180502, 2006)
which is based on many-boson quantum walk.

\vspace{.1in}

\end{abstract}

%%%%%%%%%%%%%%%%%%%%%%%%%%%%%%%%%%%%%%%%%%%%%%%%%%%%%%%%%%%%%%%%%%%%%%%%%%%%%%%%%%%%%%%%%%%%%%%%%%%%%%%%%%%%%%%%%%

\section{Introduction}

Perfect state transfer in continuous-time quantum walk on graphs has received considerable 
attention in quantum information and computation. This is in large part due to its potential 
applications in quantum information transmission over networks and its role in quantum computation.
Recently, continuing ideas developed by Childs \cite{childs09}, Underwood and Feder \cite{uf10} used 
perfect state transfer to show that continuous-time quantum walk is a universal computational model.
The notion of perfect state transfer was introduced by Bose \cite{bose03} in the context of 
information transfer on linear spin-chains. His original scheme can be generalized to arbitrary 
graphs (as described by Albanese \etal and Christandl \etal \cite{acde04,cdel04,cddekl05})
which we briefly outline in the following.

Given a weighted graph $G$ on $n$ vertices with adjacency matrix $A(G)$,
we imagine a collection of $n$ qubits associated with each vertex of $G$ and arranged so that their 
interaction is governed by a Hamiltonian $\HS_{G}$ which depends on the edge structure of $G$.
Here, our collective Hilbert space is $\bigotimes_{u \in V} \CC^{2}$ which is $2^{n}$-dimensional.
Suppose an arbitrary one-qubit state $\ket{\psi} = \alpha_{0}\ket{0} + \alpha_{1}\ket{1}$ 
is located at vertex $a$ of $G$. Our goal is to move this state to vertex $b$.
For simplicity, we depict $a$ as the leftmost qubit whereas $b$ is the rightmost qubit.
The initial configuration has the qubit at vertex $a$ be in state $\ket{\psi}$ and the 
other qubits are in the $\ket{0}$ state, while the final configuration has the qubit at 
vertex $b$ be in state $\ket{\psi}$ with the other qubits being in the $\ket{0}$ state:
\begin{eqnarray}
\ket{\mbox{\sc start}} 
	& = & \ket{\psi}_{a} \otimes \ket{0} \otimes \ldots \otimes \ket{0} \otimes \ket{0}_{b} 
		\ = \ \alpha_{0}\ket{00 \ldots 00} + \alpha_{1}\ket{10 \ldots 00} \\
\ket{\mbox{\sc final}}
	& = & \ket{0}_{a} \otimes \ket{0} \otimes \ldots \otimes \ket{0} \otimes \ket{\psi}_{b} 
		\ = \ \alpha_{0}\ket{00 \ldots 00} + \alpha_{1}\ket{00 \ldots 01}.
\end{eqnarray}
The main goal of perfect state transfer is to achieve, at some time $t$,
\begin{equation} \label{eqn:big-pst}
|\bra{\mbox{\sc final}}e^{-it\HS_{G}}\ket{\mbox{\sc start}}| = 1.
\end{equation}
Natural assumptions can be placed on $\HS_{G}$ which will allow us to view (\ref{eqn:big-pst}) 
as a continuous-time quantum walk on $G$.

For example, one typically assumes $\HS_{G}$ {\em commutes} with $\ZS = \sum_{u \in V} Z_{u}$, where 
the latter operator counts the number of qubits in the $\ket{1}$ state\footnote{Here, 
$Z_{u}$ denotes an $n$-fold tensor product of identity matrices except at position $u$ which has the 
Pauli $Z$ matrix; the same convention applies to the other Pauli matrices.}.
%$[\HS_{G},\ZS] = 0$.
Note that the eigenvalues of $\ZS$ are given by $\lambda_{k} = -n + 2k$, for $k=0,1,\ldots,n$.
%$\lambda_{0}=-n$, $\lambda_{1}=-n+2$, $\ldots$, $\lambda_{n-1}=n-2$, $\lambda_{n}=n$.
%This is because each $Z_{w}$ commutes with each term $X_{u}X_{v} + Y_{u}Y_{v}$ of $H_{G}$.
Since $\ket{00 \ldots 00}$ belongs to the zero eigenspace of $\HS_{G}$,
%and thus remains unchanged under the unitary evolution $\exp(-it\HS_{G})$.
we may focus on the unitary evolution of $\ket{10 \ldots 00}$ under $e^{-it\HS_{G}}$.
The latter state belongs to the eigenspace $\Lambda_{1}$ of $\ZS$ corresponding to the eigenvalue 
$\lambda_{1} = -n+2$ (since it has exactly one qubit in the $\ket{1}$ state). Thus,
\begin{equation}
e^{-it\HS_{G}}\left(\alpha_{0}\ket{00 \ldots 00} + \alpha_{1}\ket{10 \ldots 00}\right)
\ = \
\alpha_{0}\ket{00 \ldots 00} + \alpha_{1} e^{-it\HS_{G}}\ket{10 \ldots 00}.
\end{equation}
Since $\HS_{G}$ and $\ZS$ commute, the eigenspace $\Lambda_{1}$ is $\HS_{G}$-invariant;
this is because if $\ket{z}$ is an eigenvector of $\ZS$ with eigenvalue $\lambda$, 
then so is $\HS_{G}\ket{z}$. 
%This follows from the commutativity relation $\ZS \HS_{G}\ket{z} = \HS_{G}\ZS\ket{z} = \lambda \HS_{G}\ket{z}$.
Thus, the time evolution of $\exp(-it\HS_{G})\ket{10 \ldots 00}$ stays inside the eigenspace $\Lambda_{1}$.
Moreover, we have the following basis states for $\Lambda_{1}$:
\begin{equation}
\ket{1} = \ket{100 \ldots 0}, \ \ \
\ket{2} = \ket{010 \ldots 0}, \ \ \
\ldots, \ \ \
\ket{n} = \ket{000 \ldots 1}
\end{equation}
which forms a natural correspondence with the vertices of $G$;
thus, $\ket{a} = \ket{1}$ and $\ket{b} = \ket{n}$.
Furthermore, suppose $\HS_{G}$ agrees with $A(G)$ on the subspace $\Lambda_{1}$
where $\bra{v}\HS_{G}\ket{u}$ equals the weight $\omega_{u,v}$ of the edge $(u,v)$, for all $u,v \in V$.
Examples of $\HS_{G}$ satisfying these assumptions include the XY exchange Hamiltonian
$\HS_{G} = \frac{1}{2} \sum_{(u,v) \in E(G)} \omega_{u,v} (X_{u}X_{v} + Y_{u}Y_{v})$,
as well as the Heisenberg exchange (which is related to the Laplacian of $G$).
This shows that the $2^n$-dimensional time evolution $e^{-it\HS_{G}}\ket{\mbox{\sc start}}$
can be viewed as a $n$-dimensional time evolution $e^{-itA(G)}\ket{a}$ since the former is confined
to the single-excitation subspace $\Lambda_{1}$. 
Further background on these connections may be found in \cite{bose03,cddekl05,o06,agrr07}.

By the preceding arguments, we may study perfect state tranfer (\ref{eqn:big-pst}) 
as a {\em continuous-time quantum walk} on the graph $G$ (see Farhi and Gutmann \cite{fg98}).
Thus, without loss of generality, we say a graph $G=(V,E)$ has {\em perfect state transfer} (PST) 
from $a$ to $b$ at time $t$ if
\begin{equation} \label{eqn:pst-simplified}
|\bra{b}e^{-itA(G)}\ket{a}|=1,
\end{equation}
where $A(G)$ is the adjacency matrix of $G$ (thus, focusing on the XY interaction model).
This allows us to investigate mathematical properties of the graph $G$ which enable such phenomena to occur.
The reduction to quantum walk on graphs was a crucial element in the early works on perfect state transfer 
(see \cite{bose03,acde04,cdel04,cddekl05}).

Christandl \etal \cite{cdel04,cddekl05} showed that taking an $k$-fold Cartesian graph product of either
a $2$-path or a $3$-path (that is, $K_{2}$ or $P_{3}$) with itself yields a high-diameter graph which has 
perfect state transfer. This follows since both $K_{2}$ and $P_{3}$ have antipodal perfect state transfer 
and because the Cartesian product operation preserves perfect state transfer.
They also made a crucial connection between hypercubes and weighted paths using the so-called 
{\em path-collapsing} argument\footnote{This argument was used earlier by Childs \etal \cite{ccdfgs03} in 
the context of exponential algorithmic speedup for a graph search problem.}. 
Christandl \etal \cite{cddekl05} also observed that the $n$-path $P_{n}$, for $n \ge 4$, has no
antipodal perfect state transfer but a suitably weighted version of it has perfect state transfer.
The weighting scheme on $P_{n}$ is derived from a path-collapsed $(n-1)$-cube. 

In an intriguing work, Feder \cite{f06} generalized the construction of the weighted paths with 
perfect state transfer described by Christandl \etal in \cite{cdel04,cddekl05}. His construction 
used a many-boson quantum walk on a single primary graph. He showed that this induced a single-boson 
quantum walk on a secondary graph and that the secondary graph has perfect state transfer if the 
primary graph does. In this construction, the weighted path of length $n$ is obtained from a 
$(n-1)$-boson quantum walk on $K_{2}$.

In algebraic graph theory, the main question is to find a characterization of graphs which exhibit 
perfect state transfer. 
Some progress on specific families of graphs were given by Bernasconi \etal \cite{bgs08} and 
by Cheung and Godsil \cite{cg11} for hypercubic graphs and by Ba\v{s}i\'{c} and Petkovi\'{c} \cite{bp09} 
for circulant graphs. Although a general characterization remains beyond the reach of current methods,
strong general results towards this goal were recently proved by Godsil \cite{godsil-survey,godsil-pst}.
In one of his results, Godsil proved that a necessary condition for $G$ to have perfect state transfer 
between vertices $a$ and $b$ is that they are {\em cospectral}, that is, the vertex-deleted subgraphs 
$G \setminus a$ and $G \setminus b$ are isomorphic. This intuitively suggests that the neighborhoods
around $a$ and $b$ must look similar. In fact, prior to this work, all known examples of graphs with 
perfect state transfer between vertices $a$ and $b$ admit an automorphism which maps $a$ to $b$. 
In \cite{kay11}, Kay proved that the latter property is necessary for paths, while
in \cite{godsil-survey}, Godsil asked if this necessary condition holds for any graph.

%%%%%%%%%%%%%%%%%%%%%%%%%%%%%%%%%%%%%%%%%%%%%%%%%%%%%%%%%%%%%%
%% K2(Box 3) = Q3 and P3(Box 2)
%% Christandl et al.
%%%%%%%%%%%%%%%%%%%%%%%%%%%%%%%%%%%%%%%%%%%%%%%%%%%%%%%%%%%%%%
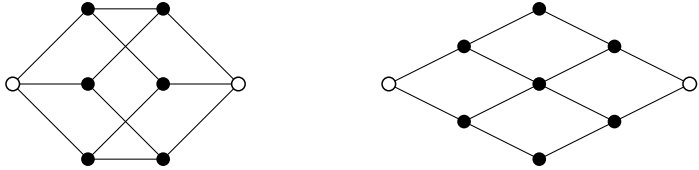
\begin{figure}[t]
\begin{center}
\begin{tikzpicture}
\draw (1,1)--(0,0)--(1,-1);
\draw (0,0)--(1,0);
\draw (1,1)--(2,0)--(1,-1);
\draw (2,1)--(1,0)--(2,-1);
\draw (2,1)--(3,0)--(2,-1);
\draw (2,0)--(3,0);
\draw (1,1)--(2,1);
\draw (1,-1)--(2,-1);

\foreach \x in {1,2}
\foreach \y in {-1,0,1}
	\node at (\x,\y)[circle, fill=black][scale=0.5]{};
% pst nodes
\foreach \x in {0,3}
{
	\node at (\x, 0)[circle, fill=white][scale=0.5]{};
	\draw[line width=0.2mm] (\x,0) circle (0.09cm);
}

\draw (5,0)--(6,0.5)--(7,1)--(8,0.5)--(9,0)--(8,-0.5)--(7,-1)--(6,-.5)--(5,0);
\draw (6,0.5)--(7,0)--(8,-0.5);
\draw(6,-0.5)--(7,0)--(8,0.5);
\foreach \y in {-1,0,1}
	\node at (7, \y)[circle, fill=black][scale=0.5]{};
\foreach \x in {6,8} 
	\foreach \y in {-0.5,0.5} 
		\node at (\x, \y)[circle, fill=black][scale=0.5]{};
% pst nodes
\foreach \x in {5,9} 
{
	\node at (\x, 0)[circle, fill=white][scale=0.5]{};
	\draw[line width=0.2mm] (\x,0) circle (0.09cm);
}
\end{tikzpicture}
\vspace{.1in}
\caption{The Cartesian product construction for perfect state transfer (PST): 
(a) $K_{2}^{\ecart 3}$; 
(b) $P_{3}^{\ecart 2}$ (see Christandl \etal \cite{cddekl05}).
Antipodal PST occurs between vertices marked white.
}
\vspace{.1in}
\label{fig:cartesian}
\end{center}
\hrule
\end{figure}

Our goal in the present work is to explore the role of quotient graphs in perfect state transfer.
Since quotient graphs naturally arise in the context of equitable partitions, we use this
formalization to capture the idea behind path-collapsing arguments \cite{cdel04,cddekl05,ccdfgs03}.
We argue that equitable partitions provide the most natural way to view these arguments since the 
resulting proofs are more transparent. Moreover, equitable partitions have been studied 
extensively in algebraic graph theory (see Godsil and McKay \cite{gm80} and 
Godsil and Royle \cite{godsil-royle01}) and have a well-established collection of results 
which we can build upon.

A main observation we use throughout is the following statement which admits a simple proof:
a graph $G$ has perfect state transfer if and only if its quotient graph $G/\pi$ modulo an 
equitable partition $\pi$ has perfect state transfer. 
Although {\em weaker} forms of this statement had appeared in different guises before, 
we give a simple and direct proof using the machinery of equitable partitions.
The necessary condition was used by Christandl \etal \cite{cdel04,cddekl05} to establish that 
certain weighted paths have perfect state transfer (in contrast to its unweighted variants).
Childs \etal \cite{ccdfgs03} used the sufficient condition to analyze hitting times of specific
graphs related to binary trees.
We will use the backward implication of the equivalence to construct new perfect state transfer graphs.
In our first application, we use this {\em lifting} property to construct a graph with perfect state 
transfer between two vertices but has no automorphism which maps one vertex to the other. 
This answers the aforementioned question posed by Godsil \cite{godsil-survey}.

Using equitable partitions, we also provide an algebraic framework to Feder's construction.
We prove that the secondary graph obtained from a $k$-boson quantum walk on a primary graph
$G$ is equivalent to a quotient of the $k$-fold Cartesian product of $G$, that is, $G^{\ecart k}/\pi$, 
for some equitable partition $\pi$. This equivalence is related to works 
by Audenaart \etal \cite{agrr07} on symmetric powers of graphs 
and by Osborne \cite{o06} on wedge product on graphs.
Our work differs from \cite{agrr07} in that we preserve diagonal entries and from \cite{o06} in
that we work in a symmetric vector space (rather than {\em exterior} vector space).
A common thread in all these works is the use of algebraic graph theory to provide 
an explicit connection between many-particle and single-particle quantum walks.
Another related work along the same lines was given in \cite{wm09}.
In our algebraic formalism, we employ a model of many-particle quantum walk used 
by Gamble \etal \cite{gfzjc10} and by Smith \cite{smith10} in their works on graph 
isomorphism.

%%%%%%%%%%%%%%%%%%%%%%%%%%%%%%%%%%%%%%%%%%%%%%%%%%%%%%%%%%%%%%
%% title: a sequence of Feder graphs P3 with 2,3, and 4 bosons
%% author: Michael Landry, June/July 2011
%%%%%%%%%%%%%%%%%%%%%%%%%%%%%%%%%%%%%%%%%%%%%%%%%%%%%%%%%%%%%%
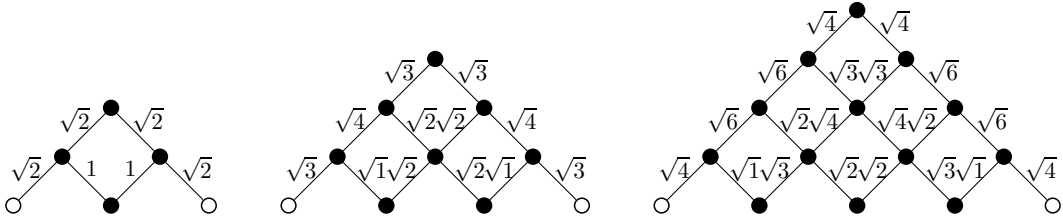
\begin{figure}[t]
\begin{center}
\begin{tikzpicture}[scale=0.65]

\draw(-2,0)--(0,2);
\draw(0,0)--(1,1);
\draw(0,2)--(2,0);
\draw(-1,1)--(0,0);

% pst nodes
\foreach \x in {-2,2}
{
	\node at (\x, 0)[circle, fill=white][scale=0.6]{};
	\draw[line width=0.2mm] (\x, 0) circle (0.15cm);
}

\node at (0,2)[circle, fill=black][scale=0.6]{};
\node at (0,0)[circle, fill=black][scale=0.6]{};
\node at (1,1)[circle, fill=black][scale=0.6]{};
\node at (-1,1)[circle, fill=black][scale=0.6]{};

\node at (-1.75,.75)[scale=0.8]{$\sqrt2$};
\node at (-.75,1.75)[scale=0.8]{$\sqrt2$};
\node at (-.4,.75)[scale=0.8]{1};
\node at (.4,.75)[scale=0.8]{1};
\node at (.75,1.75)[scale=0.8]{$\sqrt2$};
\node at (1.75,.75)[scale=0.8]{$\sqrt2$};
\end{tikzpicture}\quad \quad
%%%%%%%%%%%%%%%%%%%%%%%%%%%%%%%%%%%%%%%%%%%%%
\begin{tikzpicture}[scale=0.65]
%edges
\draw(-3,0)--(0,3);
\draw(-1,0)--(1,2);
\draw(1,0)--(2,1);
\draw(3,0)--(0,3);
\draw(1,0)--(-1,2);
\draw(-1,0)--(-2,1);

%vertices
% pst nodes
\foreach \a in {-3, 3}
{
	\node at (\a, 0)[circle, fill=white][scale=0.6]{};
	\draw[line width=0.2mm] (\a, 0) circle (0.15cm);
}
\foreach \a in {-1,1}
\node at (\a,0)[circle, fill=black][scale=0.6]{};
\foreach \b in {-2,0,2}
\node at (\b,1)[circle, fill=black][scale=0.6]{};
\foreach \c in {-1,1}
\node at (\c,2)[circle, fill=black][scale=0.6]{};
\node at (0,3)[circle, fill=black][scale=0.6]{};

%edgeweights

\foreach \a in {-2.75,2.75}
\node at (\a, 0.75)[scale=0.8]{$\sqrt3$};
\foreach \b in {-1.75,1.75}
\node at (\b, 1.75)[scale=0.8]{$\sqrt4$};
\foreach \c in {-.75,.75}
\node at (\c, 2.75)[scale=0.8]{$\sqrt3$};
\foreach \d in {-.7,0.7}
\node at (\d,0.75)[scale=0.8]{$\sqrt2$};
\foreach \e in {-.3,.3}
\node at (\e, 1.75)[scale=0.8]{$\sqrt2$};
\foreach \f in {-1.3,1.3}
\node at (\f,0.75)[scale=0.8]{$\sqrt1$};
\end{tikzpicture}\quad \quad
%%%%%%%%%%%%%%%%%%%%%%%%%%%%%%%%%%%%%%%%%%%%%%
\begin{tikzpicture}[scale=0.65]
%edges
\draw (-4,0)--(0,4);
\draw (0,4)--(4,0);
\draw (-2,0)--(1,3);
\draw(-1,3)--(2,0);
\draw(-3,1)--(-2,0);
\draw(-2,2)--(0,0);
\draw(3,1)--(2,0);
\draw(2,2)--(0,0);

%vertices
% pst nodes
\foreach \a in {-4, 4}
{
	\node at (\a, 0)[circle, fill=white][scale=0.6]{};
	\draw[line width=0.2mm] (\a, 0) circle (0.15cm);
}
\foreach \a in {-2,0,2}
\node at (\a, 0)[circle, fill=black][scale=0.6]{};
\foreach \b in {-3,-1,1,3}
\node at (\b,1)[circle, fill=black][scale=0.6]{};
\foreach \c in{-2,0,2}
\node at (\c,2)[circle, fill=black][scale=0.6]{};
\foreach \d in {-1,1}
\node at (\d,3)[circle, fill=black][scale=0.6]{};
\node at (0,4)[circle, fill=black][scale=0.6]{};

%edgeweights
\foreach \a in {-2.3, 2.3}
\node at (\a,0.75)[scale=0.8]{$\sqrt1$};

\foreach \b in {-1.3,1.3}
\node at (\b,1.75)[scale=0.8]{$\sqrt2$};

\foreach \c in {-.3, .3}
\node at (\c, 2.75)[scale=0.8]{$\sqrt3$};

\foreach \d in {-.75, .75}
\node at (\d,3.75)[scale=0.8]{$\sqrt4$};

\foreach \e in {-.3,.3}
\node at (\e,0.75)[scale=0.8]{$\sqrt2$};

\foreach \f in {-0.7,.7}
\node at (\f,1.75)[scale=0.8]{$\sqrt4$};

\foreach \f in {-1.75,1.75}
\node at (\f,2.75)[scale=0.8]{$\sqrt6$};

\foreach \g in {-1.7,1.7}
\node at (\g,0.75)[scale=0.8]{$\sqrt3$};

\foreach \h in {-2.75,2.75}
\node at (\h, 1.75)[scale=0.8]{$\sqrt6$};

\foreach \h in {-3.75,3.75}
\node at (\h, .75)[scale=0.8]{$\sqrt4$};
\end{tikzpicture}

\vspace{.1in}
\caption{Feder's weighted lattice PST graphs obtained from $k$-boson quantum walks on 
$P_{3}$ with $k=2,3,4$ (see Feder \cite{f06}). Equivalently, these are the quotient graphs 
$P_3^{\square k}/\pi$ (see Theorem \ref{thm:feder}).
Antipodal PST occurs between vertices marked white.
}
\vspace{.1in}
\end{center}
\hrule
\end{figure}

Finally, we explore Feder's construction when distinct primary graphs with
commensurable perfect state transfer (or even periodic) times are used.
We prove a composition theorem which shows partial commutativity between the Cartesian 
product and quotient operators.
This mixed construction is akin to perfect state transfer graphs obtained using weak and 
lexicographic products (see Ge \etal \cite{ggpt11}) and graph joins 
(see Angeles-Canul \etal \cite{anoprt10}).
We found new families of perfect state transfer graphs using cube-like graphs (which were studied
by Bernasconi \etal \cite{bgs08} and by Cheung and Godsil \cite{cg11}). 
%The first family of cube-like graphs have perfect state transfer at time $\pi/2$ whereas the second 
%family of graphs have perfect state transfer at time $\pi/4$ (and hence are periodic at time $\pi/2$).
The graphs derived from these cube-like graphs are different from weighted graphs 
obtained in Feder's construction.

Our proofs rely on basic ideas from algebraic graph theory and exploit spectral properties
of the underlying graphs.

%%%%%%%%%%%%%%%%%%%%%%%%%%%%%%%%%%%%%%%%%%%%%%%%%%%%%%%%%%%%%%%%%%%%%%%%%%%%%%%%%%%%%%%%%%%%%%%%%%%%%%%%%%%%%%%%%%

\section{Preliminaries}

For a logical statement $\mathcal{S}$, the expression $\iverson{\mathcal{S}}$ equals
$1$ if $\mathcal{S}$ is true and $0$ otherwise.
We use $[n]$ to denote $\{1,2,\ldots,n\}$. 
The all-one $m \times n$ matrix is denoted by $J_{m,n}$; 
we also use $\mathbf{j}_{n}$ to denote the all-one $n$-dimensional column vector.

The graph $G=(V,E)$ we study are finite, undirected, and connected.
The adjacency matrix $A(G)$ of $G$ is defined as $A(G)_{u,v} = \iverson{(u,v) \in E}$.
A graph $G$ is called $k$-regular if each vertex $u$ of $G$ has exactly $k$ adjacent neighbors.
We say a graph $G$ is $(n,k)$-regular if it has $n$ vertices and is $k$-regular.
%For integers $n \ge 1$ and $0 \le k < n$, let $\REG_{n,k}$ be the set of all $n$-vertex $k$-regular graphs.
The distance $d(a,b)$ between vertices $a$ and $b$ is the length of the shortest path connecting them.
For weighted graphs $G=(V,E,\omega)$, where $\omega: E \rightarrow \RR^{+}$ is the weight
function on edges, we let $A(G)_{u,v} = \omega(u,v)$ be the edge weight of $(u,v)$.

An {\em automorphism} $\tau$ of a graph $G=(V,E)$ is a bijective map on the vertex set $V$ that
respects the edge relation $E$; that is, $(u,v) \in E$ if and only if $(\tau(u),\tau(v)) \in E$.
If $P$ is a permutation matrix which represents an automorphism $\tau$ of $G$, 
then $P$ commutes with $A(G)$, or $PA(G) = A(G)P$. The automorphism group of $G$ is denoted $\Aut(G)$.

Standard graphs we consider include complete graphs $K_{n}$, paths $P_{n}$, and Cayley graphs.
For a given group $\GG$ and a subset $S \subseteq \GG$, the Cayley graph $X(\GG,S)$ has the group $\GG$
as its vertex set where two group elements $g$ and $h$ are adjacent if $gh^{-1} \in S$.
For $X(\GG,S)$ to be connected, we require $S$ to be a generating set of $\GG$. %, that is, $\lip S\rip = \GG$.
If $S$ is closed under taking inverses, that is, $S^{-1} = S$, then $X(\GG,S)$ is undirected.
An $n$-vertex {\em circulant} graph $G$ is a 
Cayley graph $X(\ZZ_{n},S)$ of the cyclic group of order $n$.
Known examples of circulants include complete graphs $K_{n}$ and cycles $C_{n}$.
%Alternatively, we may define a circulant graph $G$ on $[n]$ through a subset $S \subseteq [n]$ 
%where $j \sim k$ if and only if $j-k \in S$; we denote such a circulant as $\Circ(n,S)$.

%Let $G$ and $H$ be two graphs with adjacency matrices $A(G)$ and $A(H)$, respectively.
The complement of a graph $G=(V,E)$, denoted $\overline{G}$, is a graph where $u$ is adjacent to $v$ 
if and only if $(u,v) \not\in E$, for $u \neq v$. 
The {\em Cartesian product} $G \cart H$ is a graph defined on the vertex set $V(G) \times V(H)$
where $(g_{1},h_{1})$ is adjacent to $(g_{2},h_{2})$ if either
$g_{1}=g_{2}$ and $(h_{1},h_{2}) \in E_{H}$, or $(g_{1},g_{2}) \in E_{G}$ and $h_{1}=h_{2}$.
The adjacency matrix of $G \cart H$ is $A(G) \otimes I + I \otimes A(H)$. 
The $n$-dimensional hypercube (or $n$-cube) $Q_{n}$ is defined recursively as 
$Q_{1} = K_{2}$ and $Q_{n} = K_{2} \cart Q_{n-1}$, for $n \ge 2$.
Note $Q_{n}$ is simply the Cayley graph $X(\ZZ_{2}^{n},S)$, 
where $S$ is the standard generating set for $\ZZ_{2}^{n}$.
% (which are the unit vectors with Hamming weight one).

The {\em join} $G + H$ is a graph defined on $V(G) \cup V(H)$ obtained by taking two disjoint copies
of $G$ and $H$ and by connecting all vertices of $G$ to all vertices of $H$.
%The adjacency matrix of $G+H$ is $\begin{bmatrix} A_{G} & J \\ J & A_{H} \end{bmatrix}$.
%We assume appropriate dimensions on the identity $I$ and all-one $J$ matrices used above.
The {\em cone} of a graph $G$ is defined as $K_{1} + G$ whereas 
the {\em double cone} of $G$ is given by $\overline{K}_{2} + G$.

A (vertex) partition $\pi$ of a graph $G=(V,E)$ given by $V = \biguplus_{j=1}^{m} V_{j}$ is called 
{\em equitable} if the number of neighbors in $V_{k}$ of any vertex in $V_{j}$ is a constant $d_{j,k}$, 
independent of the choice of that vertex (see \cite{godsil-royle01}). 
We call each component $V_{j}$ a {\em partition} or a {\em cell} of $\pi$.
We say a graph $G$ has an equitable {\em distance} partition $\pi$ with respect to vertices $a$ and $b$ 
if both $a$ and $b$ belong to singleton cells.
Further background on algebraic graph theory may be found in the standard monographs by
Biggs \cite{biggs} and by Godsil and Royle \cite{godsil-royle01}.

\paragraph{Continuous-time quantum walk}
For a graph $G=(V,E)$ with adjacency matrix $A(G)$, a continuous-time quantum walk on $G$ is
defined through the time-dependent unitary matrix
\begin{equation}
U_{G}(t) = \exp(-it A(G)).
\end{equation}
This model was introduced by Farhi and Gutmann \cite{fg98}.
We say that $G$ has {\em perfect state transfer} (PST) from vertex $a$ to vertex $b$ at time $t$ if 
\begin{equation} \label{eqn:pst}
|\bra{b}U_{G}(t)\ket{a}| = 1,
\end{equation}
where $\ket{a}$, $\ket{b}$ denote the unit vectors corresponding to the vertices $a$ and $b$ of $G$,
respectively. The graph $G$ has perfect state transfer if there exist vertices $a$ and $b$ in $G$
and time $t$ for which Equation (\ref{eqn:pst}) is true.
We call a graph $G$ {\em periodic} at vertex $a$ if it has perfect state transfer from $a$ to itself
at some time $t > 0$.
Further background on quantum walks and perfect state transfer may be found in the surveys 
\cite{kempe03,k06} and \cite{kay11,godsil-survey,stevanovic11,kt11}.

%%%%%%%%%%%%%%%%%%%%%%%%%%%%%%%%%%%%%%%%%%%%%%%%%%%%%%%%%%%%%%%%%%%%%%%%%%%%%%%%%%%%%%%%%%%%%%%%%%%%%%%%%%%%%%%%%%

\section{Equitable partitions and quotient graphs}

Christandl \etal \cite{cddekl05} showed that certain weighted paths have perfect state transfer
by appealing to a {\em path-collapsing} argument. 
Their argument is based on the fact that the $n$-cube $Q_{n}$ has perfect state transfer and 
it can be {\em collapsed} to a weighted path. So they deduce that weighted paths have perfect
state transfer since the underlying $n$-cube $Q_{n}$ has this property.
This argument was used in the opposite direction 
by Childs \etal \cite{ccdfgs03} in the context of exponential algorithmic speedup of a quantum
walk search algorithm. Here, they deduced properties of the underlying unweighted graphs based
on properties of the weighted paths.

A natural way to view this path-collapsing argument is via equitable partitions. 
The benefit of this is evident in the simple algebraic equivalence of perfect state transfer 
between a graph and its quotient.
The notion of {\em equitable partition} was introduced by Godsil and McKay \cite{gm80}
in their work on walk-regular graphs. Our treatment here follows closely the ones given 
by Godsil and Royle \cite{godsil-royle01} and by Godsil \cite{godsil-pst,godsil-survey}.

Let $G=(V,E)$ be a graph with an equitable partition $\pi = \biguplus_{k=1}^{m} V_{k}$ into $m$ cells.
For each $j,k \in [m]$, let $d_{j,k}$ be the number of neighbors in $V_{k}$ of any vertex 
in $V_{j}$ (which is independent of the choice of the vertex).
The {\em partition matrix} $P$ associated with $\pi$ is defined as the $|V| \times m$ matrix
where $P_{x,k}$ equals $1$ if vertex $x$ belongs to partition $V_{k}$, and equals $0$ otherwise;
that is, $P_{x,k} = \iverson{x \in V_{k}}$. 
The quotient graph $G/\pi$ defined in the literature is a weighted directed graph
whose adjacency matrix is defined as $B(G/\pi)_{j,k} = d_{j,k}$.
A fundamental fact here is that $A(G)P = PB(G/\pi)$ 
(see \cite{godsil-royle01}, Lemma 9.3.1, page 196).
 
%%%%%%%%%%%%%%%%%%%%%%%%%%%%%%%%%%%%%%%%%%%%%%%%%%%%%%%%%%%%%%%%%%%%%%%%%%%%%%%%%%%%%%%%%%%%%%%%%%%%%%%%

%%%%%%%%%%%%%%%%%%%%%%%%%%%%%%%%%%%%%%%%%%%%%%%%%%%%%%%%%%%%%%
%% cube-like graph (source: Bernasconi, Godsil, Severini)
%% and its quotient (plus its Box representation)
%%%%%%%%%%%%%%%%%%%%%%%%%%%%%%%%%%%%%%%%%%%%%%%%%%%%%%%%%%%%%%
\begin{figure}[t]
\begin{center}
\begin{tikzpicture}
% 3-cube BGS
\draw (1,1)--(0,0)--(1,-1);
\draw (0,0)--(1,0);
\draw (1,1)--(2,0)--(1,-1);
\draw (2,1)--(1,0)--(2,-1);
\draw (2,1)--(3,0)--(2,-1);
\draw (2,0)--(3,0);
\draw (1,1)--(2,1);
\draw (1,-1)--(2,-1);
\draw (0,0)--(2,1);
\draw (1,-1)--(3,0);
\draw (1,0)--(1,1);
\draw (2,0)--(2,-1);

\foreach \x in {1,2}
\foreach \y in {-1,0,1}
	\node at (\x,\y)[circle, fill=black][scale=0.5]{};
\node at (3, 0)[circle, fill=black][scale=0.5]{};

% pst nodes
\node at (0, 0)[circle, fill=white][scale=0.5]{};
\draw[line width=0.2mm] (0, 0) circle (0.09cm);
\node at (1, -1)[circle, fill=white][scale=0.5]{};
\draw[line width=0.2mm] (1, -1) circle (0.09cm);

% quotient arrow
\draw[>=latex,->](4,0)--(5,0);
\node at (4.5,0.25){$\pi$};

% collapsed 3-cube BGS
\draw (6,0)--(7,1)--(8,0)--(7,-1)--(6,0);

\node at (8, 0)[circle, fill=black][scale=0.5]{};
\node at (7, 1)[circle, fill=black][scale=0.5]{};

% pst nodes
\node at (6, 0)[circle, fill=white][scale=0.5]{};
\draw[line width=0.2mm] (6, 0) circle (0.09cm);
\node at (7, -1)[circle, fill=white][scale=0.5]{};
\draw[line width=0.2mm] (7, -1) circle (0.09cm);

%\draw (1,0) .. controls (0.5,0.5) and (0.5,1.0) .. (1,1) .. controls (1.5,1.0) and (1.5,0.5) .. (1,0);
\draw (7,1) .. controls (6.5,1.25) and (6.5,1.5) .. (7,1.5) .. controls (7.5,1.5) and (7.5,1.25) .. (7,1);
\draw (8,0) .. controls (8.25,0.5) and (8.5,0.5) .. (8.5,0) .. controls (8.5,-0.5) and (8.25,-0.5) .. (8,0);

% weights on self-loops
\node at (7.0, 1.7)[scale=0.8]{$2$};
\node at (8.65, 0)[scale=0.8]{$2$};

% other non-unit weights 
\node at (6.25, 0.75)[scale=0.8]{$\sqrt{3}$};
\node at (7.75, -0.65)[scale=0.8]{$\sqrt{3}$};

% quotient arrow
%\draw[>=latex,->](4,0)--(5,0);
\node at (9.75, 0){$\cong$};

\draw (11, -0.5)--(11, 0.5);
% pst nodes
\foreach \y in {-0.5, 0.5}
{
	\node at (11, \y)[circle, fill=white][scale=0.5]{};
	\draw[line width=0.2mm] (11, \y) circle (0.09cm);
}

\node at (12, 0){$\cart$};

\draw (13, -0.5)--(13, 0.5);
%\draw (13,-0.5) .. controls (12.5,-0.75) and (12.5,-1) .. (13,-1) .. controls (13.5,-1) and (13.5,-0.75) .. (13,-0.5);
\draw (13,0.5) .. controls (12.5,0.75) and (12.5,1) .. (13,1) .. controls (13.5,1) and (13.5,0.75) .. (13,0.5);

\node at (13, 0.5)[circle, fill=black][scale=0.5]{};
% pst node
\node at (13, -0.5)[circle, fill=white][scale=0.5]{};
\draw[line width=0.2mm] (13, -0.5) circle (0.09cm);

\node at (13, 1.25)[scale=0.8]{$2$};
\node at (13.3, 0.1)[scale=0.8]{$\sqrt{3}$};

\end{tikzpicture}
\vspace{.1in}
\caption{The cube-like graph $X(\ZZ^{3},\{100,010,001,011\})$ (see Bernasconi \etal \cite{bgs08}).
Its quotient graph is a Cartesian product of a perfect state transfer and a periodic graph (both at time $\pi/2$).}
\vspace{.1in}
\end{center}
\label{fig:bgs-cube}
\hrule
\end{figure}
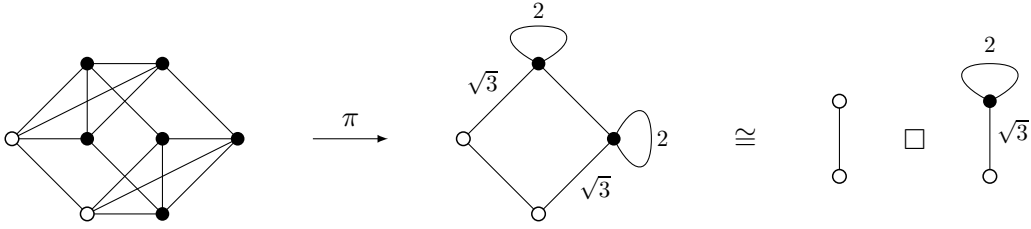

%%%%%%%%%%%%%%%%%%%%%%%%%%%%%%%%%%%%%%%%%%%%%%%%%%%%%%%%%%%%%%%%%%%%%%%%%%%%%%%%%%%%%%%%%%%%%%%%%%%%%%%%

We focus on quotient graphs which are {\em undirected}.
So, we consider the {\em normalized} partition matrix $Q$ defined as
\begin{equation}
Q = \sum_{k=1}^{m} \frac{1}{\sqrt{|V_{k}|}} P\ketbra{k}{k}.
\end{equation}
Note that $Q_{x,k} = |V_{k}|^{-1/2} P_{x,k}$, and so $Q$ is simply $P$ with each column normalized.
Moreover, we still have the fundamental relation $A(G)Q = QA(G/\pi)$, where $A(G/\pi)$ is a symmetric 
matrix defined by
\begin{equation}
A(G/\pi)_{j,k} = \sqrt{d_{j,k}d_{k,j}}.
\end{equation}
So, $A(G/\pi)$ describes a weighted graph $G/\pi$ which is an {\em undirected} quotient 
graph of $G$ with respect to the equitable partition $\pi$. 
We state further useful properties of the partition matrix $Q$.

\begin{lemma} \label{lemma:equitable-partition} (Godsil \cite{godsil-pst,godsil-survey})
The following properties on $Q$ hold:
\begin{enumerate}
\item $Q^{T}Q = I_{m}$
\item $QQ^{T} = \diag(|V_{k}|^{-1} J_{|V_{k}|})_{k=1}^{m}$.
\item $QQ^{T}$ commutes with $A(G)$.
\item $A(G/\pi) = Q^{T} A(G) Q$.
\end{enumerate}
\end{lemma}

The following theorem relates the perfect state transfer properties of a graph $G$ and its quotient
$G/\pi$ with respect to an equitable distance partition $\pi$. 
A similar statement appeared in Ge \etal \cite{ggpt11} but our proof here is simpler and more direct.

\begin{theorem} \label{thm:equal-fidelity}
Let $G=(V,E)$ be a graph with an equitable partition $\pi$ where vertices $a$ and $b$ belong to singleton cells.
Then, for any time $t$
\begin{equation}
\bra{b}e^{-itA(G)}\ket{a} = \bra{\pi(b)}e^{-itA(G/\pi)}\ket{\pi(a)}.
\end{equation}
Therefore, $G$ has perfect state transfer from $a$ to $b$ at time $t$ 
if and only if 
$G/\pi$ has perfect state transfer from $\pi(a)$ to $\pi(b)$ at time $t$.
\end{theorem}
\prf
Since $A(G)$ commutes with $QQ^{T}$, we have $(QQ^{T}A(G))^{k} = A(G)^{k}QQ^{T}$ for $k \ge 1$.
Given that $a$ and $b$ are in singleton cells, 
$\ket{\pi(a)} = Q^{T}\ket{a}$ and $\ket{\pi(b)} = Q^{T}\ket{b}$. 
Thus, we have
\begin{eqnarray}
\bra{\pi(b)}e^{-itA(G/\pi)}\ket{\pi(a)}
	& = & \bra{\pi(b)} e^{-it Q^{T}A(G)Q} \ket{\pi(a)} \\
	%\mbox{ since $A(G/\pi) = Q^{T}A(G)Q$} \\
	& = & \bra{b}Q \left[ \sum_{k=0}^{\infty} \frac{(-it)^{k}}{k!} (Q^{T}A(G)Q)^{k} \right] Q^{T}\ket{a} \\
	& = & \bra{b} \left[ \sum_{k=0}^{\infty} \frac{(-it)^{k}}{k!} (QQ^{T}A(G))^{k} \right] QQ^{T}\ket{a}, \ \
	\mbox{ by regrouping } \\
	& = & \bra{b} e^{-it A(G)} QQ^{T}\ket{a},
	%\mbox{ using $(Q^{T}QA(G))^{k}=A(G)^{k}QQ^{T}$} \\
	%& = & \bra{b} e^{-itA(G)} \ket{a},
\end{eqnarray}
which proves the claim since $QQ^{T}\ket{a} = \ket{a}$ because $a$ belongs to a singleton cell.
\qed

%%%%%%%%%%%%%%%%%%%%%%%%%%%%%%%%%%%%%%%%%%%%%%%%%%%%%%%%%%%%%%%%%%%%%%%%%%%%%%%%%%%%%%%%%%%%%%%%%%%%%%%%%%%%%%%%%%

\section{Lifting graph constructions} \label{section:xeg}

In this section, we focus on the backward implication of Theorem \ref{thm:equal-fidelity}.
This is a {\em lifting} theorem which states if a quotient graph $G/\pi_{1}$ has perfect state transfer,
for some equitable partition $\pi_{1}$, then the graph $G$ itself must have perfect state transfer. 
This also implies that any quotient of $G$, say $G/\pi_{2}$, for any other equitable partition $\pi_{2}$, 
has perfect state transfer. We use this property to construct new graphs with perfect state transfer.

In \cite{godsil-survey}, Godsil asked the following question:
{\em if a graph $G$ has perfect state transfer between vertices $a$ and $b$, does there exist 
an automorphism of $G$ which maps $a$ to $b$?} 
We contrast this to Kay's notion of a {\em symmetry} operator $S$ on $G$ which is a unitary
operator satisfying $SA(G)=A(G)S$ and $S\ket{a}=\ket{b}$. In this latter case, Kay \cite{kay11} proved 
that such an operator $S$ always exists; but Godsil's question went further and asked if there always exists 
such an $S$ which is also a graph automorphism of $G$. The question is interesting since, prior to this
work, all known graphs with perfect state transfer exhibit this automorphism property.

We answer Godsil's question in the negative by constructing a perfect state transfer graph which lacks 
the requisite automorphism.
Our construction proceeds by lifting a simple weighted $4$-vertex path onto a glued double-cone graph. 
The latter graph was considered earlier in Ge \etal \cite{ggpt11} but in a completely different context.
We start with a simple observation.

%%%%%%%%%%%%%%%%%%%%%%%%%%%%%%%%%%%%%%%%%%%%%%%%%%%%%%%%%%%%%%
%% Small-path lifting: P4 and P5
%% source: Michael Landry & Jessica Fuller
%%%%%%%%%%%%%%%%%%%%%%%%%%%%%%%%%%%%%%%%%%%%%%%%%%%%%%%%%%%%%%
\begin{figure}[t]
\begin{center}
\begin{tikzpicture}[scale=0.9]

% P4 and its lifted graph
\foreach \z in {-8}
\foreach \y in {0}
\foreach \w in {0.25}
{
\draw(-2+\z, 0+\y)--(2+\z, 0+\y);

\draw (-1+\w+\z, 0+\y) .. controls (-1.5+\w+\z, 0.5+\y) and (-1.5+\w+\z, 0.75+\y) .. (-1+\w+\z, 0.75+\y) .. controls (-0.5+\w+\z, 0.75+\y) and (-0.5+\w+\z, 0.5+\y) .. (-1+\w+\z, 0+\y);
\draw (1-\w+\z, 0+\y) .. controls (0.5-\w+\z, 0.5+\y) and (0.5-\w+\z, 0.75+\y) .. (1-\w+\z, 0.75+\y) .. controls (1.5-\w+\z, 0.75+\y) and (1.5-\w+\z, 0.5+\y) .. (1-\w+\z, 0+\y);

\node at (-1+\w+\z, 1+\y)[scale=0.8]{$a$};
\node at (1-\w+\z, 1+\y)[scale=0.8]{$a$};

\node at (-1+\w+\z, 0+\y)[circle, fill=black][scale=0.5]{};
\node at (1-\w+\z, 0+\y)[circle, fill=black][scale=0.5]{};

% pst nodes
\foreach \x in {-2, 2}
{
	\node at (\x+\z, 0+\y)[circle, fill=white][scale=0.5]{};
	\draw[line width=0.2mm] (\x+\z, 0+\y) circle (0.09cm);
}

\node at (-1.5+\z, 0.25+\y)[scale=0.8]{$1$};
\node at (0+\z, 0.25+\y)[scale=0.8]{$b$};
\node at (1.5+\z, 0.25+\y)[scale=0.8]{$1$};

% quotient arrow
\draw[>=latex,->](0+\z, 1.5+\y)--(0+\z, 2.5+\y);
\node at (0.25+\z, 2+\y){};

\foreach \t in {0.75}
{
	% baloon diagram
	\draw[line width=0.5mm](-2+\z, 3+\y+\t)--(-1.5+\w+\z, 3+\y+\t);
	\draw[line width=0.5mm](\w-0.5+\z, 3+\y+\t)--(-\w+0.5+\z, 3+\y+\t);
	\draw[line width=0.5mm](1.5-\w+\z, 3+\y+\t)--(2+\z, 3+\y+\t);

	\node at (-1+\w+\z, 3+\y+\t)[circle, fill=none][scale=1.05]{$G_{1}$};
	\node at (1-\w+\z, 3+\y+\t)[circle, fill=none][scale=1.05]{$G_{2}$};
	\draw[line width=0.2mm] (-1+\w+\z, 3+\y+\t) ellipse (0.5cm and 0.7cm);
	\draw[line width=0.2mm] (1-\w+\z, 3+\y+\t) ellipse (0.5cm and 0.7cm);

	% pst nodes
	\foreach \x in {-2, 2}
	{
	\node at (\x+\z, 3+\y+\t)[circle, fill=white][scale=0.5]{};
	\draw[line width=0.2mm] (\x+\z, 3+\y+\t) circle (0.09cm);
	}
}

}

%\end{tikzpicture}\quad \quad

%%%%%%%%%%%%%%%%%%%%%%%%%%%%%%%%%%%%%%%%%%%%%%%%%%%%%%%%%%%%%%
%% non-hypercubic weightings for P5 (vertical orientation)
%% source: Michael Landry & Jessica Fuller
%%%%%%%%%%%%%%%%%%%%%%%%%%%%%%%%%%%%%%%%%%%%%%%%%%%%%%%%%%%%%%

%\begin{tikzpicture}

% P5
\foreach \t in {0.5} 	% shifting PST nodes towards center
{
\draw(-3+\t,-1)--(3-\t,-1);
\foreach \x in {-3+\t, 3-\t}
{
	% pst nodes
	\node at (\x, -1)[circle, fill=white][scale=0.5]{};
	\draw[line width=0.2mm] (\x, -1) circle (0.09cm);
}
\foreach \x in {-1.5,0,1.5}
{
	\node at (\x, -1)[circle, fill=black][scale=0.5]{};
}
\node at (-2.05, -0.75)[scale=0.8]{$\sqrt{2}$};
\node at (-0.75, -0.75)[scale=0.8]{$\sqrt{\alpha}$};
\node at ( 0.75, -0.75)[scale=0.8]{$\sqrt{\alpha}$};
\node at ( 2.05, -0.75)[scale=0.8]{$\sqrt{2}$};

% quotient arrow
\draw[>=latex,->](0, -0.25)--(0, 0.25);
\node at (0.25, 0){};

% lifted P5 (weighted graph)
\foreach \z in {-0.5}
{
\draw(-3+\t,2+\z)--(-1.5,2.5+\z)--(0,2.5+\z)--(1.5,2.5+\z)--(3-\t,2+\z);
\draw(-3+\t,2+\z)--(-1.5,1.5+\z)--(0,1.5+\z)--(1.5,1.5+\z)--(3-\t,2+\z);
\foreach \x in {-3+\t,3-\t}
{
	% pst nodes
	\node at (\x, 2+\z)[circle, fill=white][scale=0.5]{};
	\draw[line width=0.2mm] (\x, 2+\z) circle (0.09cm);
}
\foreach \x in {-1.5,0,1.5}
\foreach \y in {-0.5,0.5}
{
	\node at (\x, 2+\y+\z)[circle, fill=black][scale=0.5]{};
}
\foreach \x in {-0.75, 0.75}
\foreach \y in {0.25}
{
	\node at (\x, 2.5+\y+\z)[scale=0.8]{$\sqrt{\alpha}$};
	\node at (\x, 1.5-\y+\z)[scale=0.8]{$\sqrt{\alpha}$};
}
}

% quotient arrow
\draw[>=latex,->](0,2.75)--(0, 3.25);
\node at (0.25, 3){};

% lifted P5 (unweighted graph)
\foreach \y in {-1.5}
{
\draw(-3+\t,6+\y)--(-1.5,6.5+\y);
\draw[line width=0.4mm](-1.5,6.5+\y)--(-0.5,6.5+\y);
\draw[line width=0.4mm](0.5,6.5+\y)--(1.5,6.5+\y);
\draw(1.5,6.5+\y)--(3-\t,6+\y);

\draw(-3+\t,6+\y)--(-1.5,5.5+\y);
\draw[line width=0.4mm](-1.5,5.5+\y)--(-0.5,5.5+\y);
\draw[line width=0.4mm](0.5,5.5+\y)--(1.5,5.5+\y);
\draw(1.5,5.5+\y)--(3-\t,6+\y);
\foreach \x in {-3+\t,3-\t}
{
	% pst nodes
	\node at (\x, 6+\y)[circle, fill=white][scale=0.5]{};
	\draw[line width=0.2mm] (\x, 6+\y) circle (0.09cm);
}
\foreach \x in {-1.5,1.5}
\foreach \z in {-0.5,0.5}
{
	\node at (\x, 6+\z+\y)[circle, fill=black][scale=0.5]{};
}
\foreach \x in {-0.9, 0.9}
\foreach \z in {0.25}
{
	\node at (\x, 6.5+\z+\y)[scale=0.8]{$\alpha$};
	\node at (\x, 5.5-\z+\y)[scale=0.8]{$\alpha$};
}
\node at (0, 6.5+\y)[circle, fill=none][scale=0.8]{$G$};
\node at (0, 5.5+\y)[circle, fill=none][scale=0.8]{$G$};
\draw[line width=0.2mm] (0, 6.5+\y) ellipse (0.5cm and 0.4cm);
\draw[line width=0.2mm] (0, 5.5+\y) ellipse (0.5cm and 0.4cm);
}

}

\end{tikzpicture}
\vspace{.1in}
\caption{Lifting small PST paths:
(i) $\PS_{4}(a,b)$ and its lifted graph $K_{1}+G_{1}\circ G_{2}+K_{1}$,
where both $G_{1}$ and $G_{2}$ are $(n,a\sqrt{n})$-regular graphs and the connection 
between them is $(b\sqrt{n})$-regular; $G_{1}$ and $G_{2}$ need not be isomorphic.
Here $a = 2k^{2}/\sqrt{4k^2-1}$ and $b = 2(k^2-1)/\sqrt{4k^2-1}$, or vice versa,
with PST time $t=\pi/2$.
(ii) General weighting on $\PS_{5}(\alpha)$ and two of its lifted graphs,
where $G$ is the empty graph and $\alpha = 4k^2-1$, $k \ge 1$, with PST time $t=\pi/\sqrt{2}$. 
Note $k=1$ yields a quotient of the $4$-cube $Q_{4}$.
}
\vspace{.1in}
\label{fig:lift-paths}
\end{center}
\hrule
\end{figure}
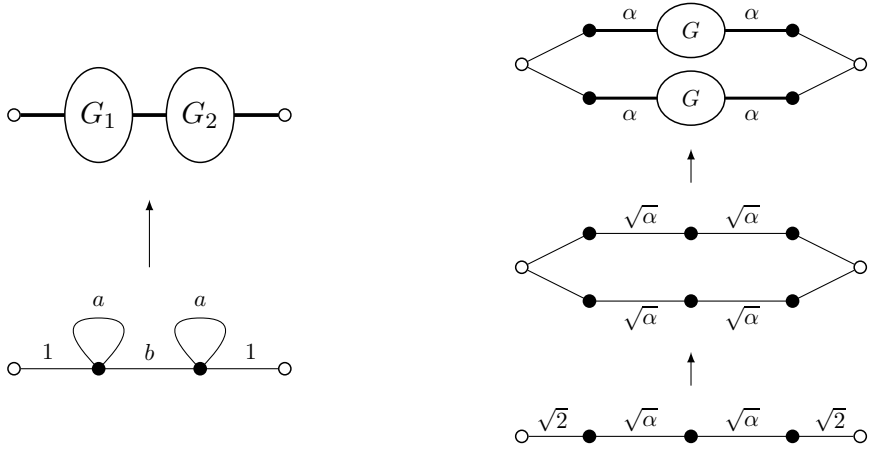

\begin{fact} \label{fact:p4}
Let $\PS_{4}(a,b)$ be a weighted path parametrized by edge-weights $a$ and $b$
(see Figure \ref{fig:lift-paths}(i))
whose adjacency matrix is:
\begin{equation}
A
=
\begin{bmatrix}
0 & 1 & 0 & 0 \\
1 & a & b & 0 \\
0 & b & a & 1 \\
0 & 0 & 1 & 0
\end{bmatrix}
\end{equation}
Let $\Delta_{\pm} = \sqrt{\frac{1}{4}(a \pm b)^{2}+1}$.
Then, $\PS_{4}(a,b)$ has antipodal (vertex $1$ to $4$) perfect state transfer at time $t$ if either
\begin{enumerate}
\item[(a)] $\cos(t\Delta_{+})\cos(t\Delta_{-}) = +1$ and $\sin(tb/2) = \pm 1$; or
\item[(b)] $\cos(t\Delta_{+})\cos(t\Delta_{-}) = -1$ and $\cos(tb/2) = \pm 1$.
\end{enumerate}
\end{fact}
\prf
Let $k_{\pm} = \frac{1}{2}(a \pm b)$ and $\Delta_{\pm}^{2} = k_{\pm}^{2} + 1$.
The eigenvalues of $\PS_{4}(a,b)$ are given by
$\alpha_{\pm} = k_{+} \pm \Delta_{+}$ and $\beta_{\pm} = k_{-} \pm \Delta_{-}$ with
the following corresponding eigenvectors
\begin{equation}
\ket{\alpha_{\pm}} = 
\frac{1}{M_{\pm}}
\begin{bmatrix}
1 & \alpha_{\pm} & \alpha_{\pm} & 1
\end{bmatrix}^{T},
\ \ \ 
\ket{\beta_{\pm}} = 
\frac{1}{N_{\pm}}
\begin{bmatrix}
1 & \beta_{\pm} & -\beta_{\pm} & -1
\end{bmatrix}^{T},
\end{equation}
where $M_{\pm}^{2} = 2(1+\alpha_{\pm}^{2})$ and $N_{\pm}^{2} = 2(1+\beta_{\pm}^{2})$
are normalization factors.
Assuming the antipodal vertices are $u$ and $v$, we have:
\begin{equation} \label{eqn:pst-p4}
\bra{v}e^{-itA}\ket{u}
=
\sum_{\pm} \frac{e^{-it\alpha_{\pm}}}{M_{\pm}^{2}}
-
\sum_{\pm} \frac{e^{-it\beta_{\pm}}}{N_{\pm}^{2}}.
\end{equation}
Since $(M_{+}M_{-})^{2} = 16\Delta_{+}^{2}$ and $(N_{+}N_{-})^{2} = 16\Delta_{-}^{2}$, we get
\begin{eqnarray} \label{eqn:expr-MN}
\sum_{\pm} \frac{e^{-it\alpha_{\pm}}}{M_{\pm}^{2}} 
& = & 
\frac{e^{-itk_{+}}}{2} \left[ \cos(t\Delta_{+}) + i\frac{k_{+}}{\Delta_{+}} \sin(t\Delta_{+})  \right] \\
\sum_{\pm} \frac{e^{-it\beta_{\pm}}}{N_{\pm}^{2}} 
& = &
\frac{e^{-itk_{-}}}{2} \left[ \cos(t\Delta_{-}) + i\frac{k_{-}}{\Delta_{-}} \sin(t\Delta_{-})  \right].
\end{eqnarray}
This proves the claim.
\qed\\

\par\noindent
The next theorem shows a construction of a family of graphs with perfect state transfer
between antipodal vertices but which has no automorphism exchanging the two vertices.

\begin{theorem} \label{thm:nonisomorphic}
For $m \ge 2$, let $n = 15 \cdot 2^{2(m-2)}$, $a = 6 \cdot 2^{m-2}$, and $b = 8 \cdot 2^{m-2}$.
Let $G_{n}$ be the family of graphs of the form $K_{1} + \AS_{n} \circ \BS_{n} + K_{1}$,
where 
$\AS_{n} = \Circ(n,\{\pm(\lfloor n/2\rfloor + 1),\ldots, \pm(\lfloor n/2\rfloor + a/2)\})$
and
$\BS_{n} = \Circ(n,\{\pm 1,\ldots,\pm a/2\})$
are two non-isomorphic families of $n$-vertex $a$-regular circulant graphs,
and the connection $\AS_{n} \circ \BS_{n}$ is given by a graph
$C_{n}$ which is an arbitrary $n$-vertex circulant of degree $b$.
Thus, the adjacency matrix of $G_{n}$ is given by:
\begin{equation}
\begin{bmatrix}
0 & \mathbf{j}_{n}^{T} & 0 & 0 \\
\mathbf{j}_{n} & \AS_{n} & C_{n} & 0 \\
0 & C_{n}^{T} & \BS_{n} & \mathbf{j}_{n} \\
0 & 0 & \mathbf{j}_{n}^{T} & 0
\end{bmatrix}
\end{equation}
Let $a_{n}$ and $b_{n}$ be the antipodal vertices of $G_{n}$.
Then $G_{n}$ has perfect state transfer between $a_{n}$ and $b_{n}$ but
there is no automorphism $\tau \in \Aut(G_{n})$ with $\tau(a_{n}) = b_{n}$.
\end{theorem}
\prf
The graph $G_{n}$ has a path-like structure with four layers where the two endpoint vertices 
have degree $n$ each and the middle two ``vertices'' are $a$-regular graphs (given by $\AS_{n}$ and
$\BS_{n}$) which are connected to each other through a $b$-regular structure (given by $C_{n}$). 
Thus, its quotient graph is a weighted $P_{4}$ whose endpoint vertices are connected by edges of weight 
$\sqrt{n}$ to the middle vertices; and, the two middle vertices have self-loops with weight $a$ each 
and are connected to each other with an edge of weight $b$; see Figure \ref{fig:lift-paths}.
After normalizing the outer two edges to unit weights, we get
$G_{n}/\pi \cong \PS_{4}(6/\sqrt{15},8/\sqrt{15})$, where $\pi$ is the equitable partition 
where the antipodal vertices belong to singleton cells.

By Fact \ref{fact:p4}, 
the quotient graph $G_{n}/\pi$ has antipodal perfect state transfer.
Therefore, we know $G_{n}$ has antipodal perfect state transfer by a lifting argument via 
Theorem \ref{thm:equal-fidelity}. It remains to show that the graphs $\AS_{n}$ and $\BS_{n}$ 
used to construct $G_{n}$ are nonisomorphic. 
This holds because $\BS_{n}$ contains too many triangles whereas $\AS_{n}$ has too few.

A triangle in a circulant $\Circ(\ZZ_{n},S)$ is given by $d_{1} + d_{2} + d_{3} \equiv 0\pmod{n}$ 
where $d_{1}$, $d_{2}$, $d_{3}$ belong to the generating set $S$.
It is clear $\BS_{n}$ has at least two triangles using $d_{1}=d_{2}=\pm 1$ and $d_{3}=\mp 2$. 
For $\AS_{n}$, we first consider the case when $m>2$ or when $n$ is even.
Each generator of $\AS_{n}$ is of the form $n/2 \pm j$, where $j \in \{1, \ldots, a/2\}$.
In this case, $d_{1}+d_{2}+d_{3} \equiv n/2 + (j_{1}+j_{2}+j_{3}) \not\equiv 0\pmod{n}$,
since $j_{1}+j_{2}+j_{3}$ is at most $3a/2$ or is at least $-3a/2$
and $3a/2 < n/2$ by the choice of $n$ and $a$.
Finally, if $m=2$, each vertex of $\AS_{n}$ is contained in exactly one triangle, by inspection.

Thus, $G_{n}$ has no automorphism which maps $a_{n}$ to $b_{n}$ (the two antipodal vertices of $G_{n}$),
since otherwise this automorphism will induce an isomorphism between $\AS_{n}$ and $\BS_{n}$. This is
because this automorphism must provide an isomorphism between the neighborhoods of $a_{n}$ and of $b_{n}$
-- which in our example are simply the graphs $\AS_{n}$ and $\BS_{n}$, respectively. This is impossible
since $\AS_{n}$ and $\BS_{n}$ are non-isomorphic.
\qed\\

%%%%%%%%%%%%%%%%%%%%%%%%%%%%%%%%%%%%%%%%%%%%%%%%%%%%%%%%%%%%%%
%% 2 non-isomorphic circulants
%% author: Michael Landry, June 2011
%%%%%%%%%%%%%%%%%%%%%%%%%%%%%%%%%%%%%%%%%%%%%%%%%%%%%%%%%%%%%%
\begin{figure}[t]
\begin{center}
\begin{tikzpicture}[scale=2.5]
\foreach \x in {18, 42,..., 360}
    \node at (\x:1)[circle,fill=black][scale=0.4] {};
\foreach \x in {114, 138,...,450}
    \foreach \y in {120, 144, 168}
    {
    \draw (\x:1)--(\x+\y:1);
    }
\node at (90:1.15) {$a_0$};
\node at (66:1.15) {$a_1$};
\node at (42:1.15) {$a_2$};
\node at (18:1.15) {$a_3$};
\node at (354:1.15) {$a_4$};
\node at (330:1.15) {$a_5$};
\node at (306:1.15) {$a_6$};
\node at (282:1.15) {$a_7$};
\node at (258:1.15) {$a_8$};
\node at (234:1.15) {$a_9$};
\node at (210:1.15) {$a_{10}$};
\node at (186:1.15) {$a_{11}$};
\node at (162:1.15) {$a_{12}$};
\node at (138:1.15) {$a_{13}$};
\node at (114:1.15) {$a_{14}$};
\end{tikzpicture}    \quad   \quad
\begin{tikzpicture}[scale=2.5]
\foreach \x in {18, 42,..., 360}
    \node at (\x:1)[circle, fill=black][scale=0.4]{};
\foreach \x in {114, 138,...,450}
    \foreach \y in {24, 48, 72}
    {
    \draw (\x:1)--(\x+\y:1);
    }
\node at (90:1.15) {$b_0$};
\node at (66:1.15) {$b_1$};
\node at (42:1.15) {$b_2$};
\node at (18:1.15) {$b_3$};
\node at (354:1.15) {$b_4$};
\node at (330:1.15) {$b_5$};
\node at (306:1.15) {$b_6$};
\node at (282:1.15) {$b_7$};
\node at (258:1.15) {$b_8$};
\node at (234:1.15) {$b_9$};
\node at (210:1.15) {$b_{10}$};
\node at (186:1.15) {$b_{11}$};
\node at (162:1.15) {$b_{12}$};
\node at (138:1.15) {$b_{13}$};
\node at (114:1.15) {$b_{14}$};
\end{tikzpicture}

\vspace{.1in}
\caption{The graphs $\mathscr{A}_2$ and $\mathscr{B}_2$ used for 
$K_{1} + \mathscr{A}_{2} \circ \mathscr{B}_{2} + K_{1}$ in Theorem \ref{thm:nonisomorphic}.}
\vspace{.1in}
\end{center}
\hrule
\end{figure}
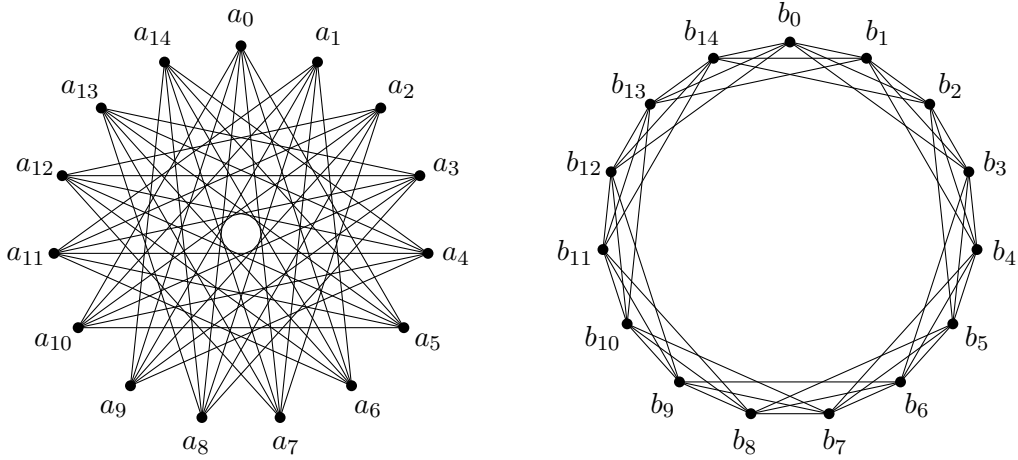

\noindent
Our lifting technique can be applied to other families of small weighted paths. 

\begin{fact} \label{fact:p5}
Let $\PS_{5}(a,b)$ be a weighted path (see Figure \ref{fig:lift-paths}(ii)) whose adjacency matrix is:
\begin{equation}
A
=
\begin{bmatrix}
0 & a & 0 & 0 & 0 \\
a & 0 & b & 0 & 0 \\
0 & b & 0 & b & 0 \\
0 & 0 & b & 0 & a \\
0 & 0 & 0 & a & 0
\end{bmatrix}
\end{equation}
Let $\Delta = a\sqrt{1 + b^2}$.
Then, $\PS_{5}(a,b)$ has antipodal perfect state transfer at time $t$ if 
$a = \sqrt{2}$, $\cos(at) = -1$ and $\cos(\Delta t)=1$.
Moreover, these conditions hold with $b = \sqrt{4k^2-1}$, for $k \ge 1$.
\end{fact}
\prf
The eigenvalues of $A$ are $0$, $\pm a$ and $\pm\Delta$ with the following corresponding eigenvectors:
\begin{eqnarray}
\ket{0} & = & \frac{1}{\sqrt{2(1+1/b^2)}} \begin{bmatrix} 1 & 0 & -a/b & 0 & 1 \end{bmatrix}^{T} \\
\ket{a_{\pm}} & = & \frac{1}{2} \begin{bmatrix} \mp 1 & -1 & 0 & +1 & \pm 1 \end{bmatrix}^{T} \\
\ket{\Delta_{\pm}} & = & \frac{1}{2\Delta/a} \begin{bmatrix} 1 & \pm\Delta/a & ab & \pm\Delta/a & 1 \end{bmatrix}^{T}
\end{eqnarray}
The choice of $a = \sqrt{2}$ is determined by the eigenvector form of $\ket{\Delta_{\pm}}$.
We leave $a$ as a variable whenever possible but use $a=\sqrt{2}$ if it leads to simpler expressions.
If the antipodal vertices are denoted $u$ and $v$, we have
\begin{equation}
\bra{v}e^{-itA}\ket{u}
=
\frac{b^2}{2(1+b^2)} - \frac{\cos(at)}{2} + \frac{\cos(\Delta t)}{2(1+b^2)}
=
\frac{b^2 + \cos(\Delta t)}{2(b^2 + 1)} - \frac{\cos(at)}{2}.
\end{equation}
To get perfect state transfer from $u$ to $v$, it suffices to require
$\cos(\Delta t)=1$ and $\cos(at)=-1$. Since $a=\sqrt{2}$, we have $t = \pi/\sqrt{2}$.
The condition $\cos(\Delta t)=1$ with $b=\sqrt{4k^2-1}$ and $t=\pi/\sqrt{2}$ is equivalent to $\cos(2\pi k)=1$,
which holds for any $k \ge 1$. 
\qed\\

\par\noindent{\em Remark}:
Fact \ref{fact:p5} 
shows that $\PS_{5}(\sqrt{2},\sqrt{4k^2-1})$, where $k \ge 1$, is a family of 
perfect state transfer paths whose first member $\PS_{5}(\sqrt{2},\sqrt{3})$ is simply the quotient 
of the cube $Q_{4} = K_{2}^{\ecart 4}$. 
Figure \ref{fig:lift-paths} shows an example of two lifted graphs obtained from this family.

%%%%%%%%%%%%%%%%%%%%%%%%%%%%%%%%%%%%%%%%%%%%%%%%%%%%%%%%%%%%%%%%%%%%%%%%%%%%%%%%%%%%%%%%%%%%%%%%%%%%%%%%%%%%%%%%%%

\section{Quotient graph constructions}

Feder \cite{f06} described an intriguing construction of perfect state transfer graphs using 
many-boson quantum walks. First, we review the basic ideas of this construction, and then we
describe its algebraic characterization using quotient graphs.

Let $G=(V,E)$ be a graph with perfect state transfer which we will call the {\em primary} graph.
For a positive integer $k$, consider a process of $k$ bosons performing a quantum walk on $G$. 
A configuration of these $k$ bosons is given by a collection of numbers $\{n_{v} : v \in V\}$, 
where $n_{v}$ represents the number of bosons located at vertex $v$, with $0 \le n_{v} \le k$.
The sum of these numbers must be $k$, that is, $\sum_{v \in V} n_{v} = k$, since there is
exactly $k$ bosons at all times.
So, a natural choice of basis states for the configurations of the $k$-boson quantum walk
is $\ket{n_{u_{1}},n_{u_{2}},\ldots,n_{u_{n}}}$, 
where $V = \{u_{1},\ldots,u_{n}\}$ is the vertex set of $G$.
The set of these basis states forms a vertex set in a so-called {\em secondary} graph.

In \cite{f06}, Feder used a nearest-neighbor hopping Hamiltonian 
$\HS = \sum_{(u,v)} a_{u}^{\dagger} a_{v}$, where $a_{u}$ and $a_{u}^{\dagger}$ are the
bosonic annihilation and creation operators.
The interaction term between the two basis states
$\ket{n_{u}, n_{v}, n_{W}}$ and $\ket{n_{u}-1, n_{v}+1, n_{W}}$
is $\sqrt{n_{u}(n_{v}+1)}$, where $n_{u} \ge 1$ and $W = V \setminus \{u,v\}$.
We summarize this construction in the following.

\begin{definition} \label{def:feder-graph} (Feder's graph \cite{f06})
Let $G=(V,E)$ be a graph and let $k \ge 1$ be a positive integer.
Let $\mathcal{G}=(\mathcal{V},\mathcal{E},\omega)$ be a weighted graph whose vertex set $\mathcal{V}$ 
is the basis states
$\ket{n_{V}} = \bigotimes \{\ket{n_{u}} : \sum_{u} n_{u} = k\}$
and whose edge set $\mathcal{E}$ is the weighted pairs
$\omega(\ket{n_{u}, n_{v}, n_{W}},\ket{n_{u}-1, n_{v}+1, n_{W}}) = \sqrt{n_{u}(n_{v}+1)}$,
assuming $n_{u} \ge 1$,
where $W = V \setminus \{u,v\}$.
We call $\mathcal{G}$ the {\em secondary} graph of $G$ with $k$ bosons, denoted by $\FF{G}{k}$. 
\end{definition}

A nice property of Feder's construction is that it generalizes the weighted paths of 
Christandl \etal \cite{cdel04,cddekl05}. As noted earlier, the latter is based on a path-collapsing 
argument of the $n$-cube. This yields a weighted path $\PS_{n+1}$ on the vertex set $\{0,1,\ldots,n\}$ 
where the edge weight of $(j,j+1)$ is $\sqrt{(j+1)(n-j)}$.
In Feder's notation, we have
$\PS_{n+1} = \FF{K_{2}}{n}$.
By recursion, this generated various infinite families of graphs with perfect state transfer,
with connections to high-dimensional Platonic solids, such as parallelepipeds, hypertetrahedra, 
hyperoctahedra (see \cite{f06}).

%%%%%%%%%%%%%%%%%%%%%%%%%%%%%%%%%%%%%%%%%%%%%%%%%%%%%%%%%%%%%%
%% K2(Box 3)=Q3 and P3(Box 2) and their quotients
%% weighted P4 and diamond D6 graphs
%%%%%%%%%%%%%%%%%%%%%%%%%%%%%%%%%%%%%%%%%%%%%%%%%%%%%%%%%%%%%%
\begin{figure}[t]
\begin{center}
\begin{tikzpicture}
% 3-cube
\draw (1,1)--(0,0)--(1,-1);
\draw (0,0)--(1,0);
\draw (1,1)--(2,0)--(1,-1);
\draw (2,1)--(1,0)--(2,-1);
\draw (2,1)--(3,0)--(2,-1);
\draw (2,0)--(3,0);
\draw (1,1)--(2,1);
\draw (1,-1)--(2,-1);

\foreach \x in {1,2}
\foreach \y in {-1,0,1}
	\node at (\x,\y)[circle, fill=black][scale=0.5]{};
% pst nodes
\foreach \x in {0,3}
{
	\node at (\x, 0)[circle, fill=white][scale=0.5]{};
	\draw[line width=0.2mm] (\x, 0) circle (0.09cm);
}

% quotient arrow
\draw[>=latex,->](1.5,-1.5)--(1.5,-2.5);
\node at (1.75,-2){$\pi$};

% collapsed p4
\draw (0,-3)--(1,-3)--(2,-3)--(3,-3);
\foreach \x in {1,2}
	\node at (\x, -3)[circle, fill=black][scale=0.5]{};
% pst nodes
\foreach \x in {0,3}
{
	\node at (\x, -3)[circle, fill=white][scale=0.5]{};
	\draw[line width=0.2mm] (\x, -3) circle (0.09cm);
}
\node at (0.5, -3.25)[scale=0.8]{$\sqrt{3}$};
\node at (1.5, -3.25)[scale=0.8]{$\sqrt{4}$};
\node at (2.5, -3.25)[scale=0.8]{$\sqrt{3}$};

% p3 box p3
\draw (5,0)--(6,0.5)--(7,1)--(8,0.5)--(9,0)--(8,-0.5)--(7,-1)--(6,-.5)--(5,0);
\draw (6,0.5)--(7,0)--(8,-0.5);
\draw(6,-0.5)--(7,0)--(8,0.5);
\foreach \y in {-1,0,1}
	\node at (7, \y)[circle, fill=black][scale=0.5]{};
\foreach \x in {6,8} 
	\foreach \y in {-0.5,0.5} 
		\node at (\x, \y)[circle, fill=black][scale=0.5]{};
% pst nodes
\foreach \x in {5,9} 
{
	\node at (\x, 0)[circle, fill=white][scale=0.5]{};
	\draw[line width=0.2mm] (\x, 0) circle (0.09cm);
}

% collapsed diamond d6
\draw (5,-3)--(6,-3)--(7,-2.5)--(8,-3)--(9,-3);
\draw (6,-3)--(7,-3.5)--(8,-3);

% pst nodes
\foreach \x in {5,9}
{
	\node at (\x, -3)[circle, fill=white][scale=0.5]{};
	\draw[line width=0.2mm] (\x, -3) circle (0.09cm);
}

\foreach \x in {6,8}
	\node at (\x, -3)[circle, fill=black][scale=0.5]{};
\foreach \y in {-2.5,-3.5}
	\node at (7, \y)[circle, fill=black][scale=0.5]{};
\node at (5.35, -3.25)[scale=0.8]{$\sqrt{2}$};
\node at (6.35, -2.35)[scale=0.8]{$\sqrt{2}$};
\node at (7.5, -2.35)[scale=0.8]{$\sqrt{2}$};
\node at (6.35, -3.5)[scale=0.8]{$1$};
\node at (7.5, -3.5)[scale=0.8]{$1$};
\node at (8.5, -3.25)[scale=0.8]{$\sqrt{2}$};

% quotient arrow
\draw[>=latex,->](7,-1.5)--(7,-2);
\node at (7.25,-1.75){$\pi$};

\end{tikzpicture}
\vspace{.1in}
\caption{The Cartesian product graphs $K_{2}^{\ecart 3}$ and $P_{3}^{\ecart 2}$ and their quotients
under equitable partitions, whose cells are orbits of $S_{3}$ and $S_{2}$ acting on the respective vertex sets.
}
\vspace{.1in}
\label{fig:quotients}
\end{center}
\hrule
\end{figure}
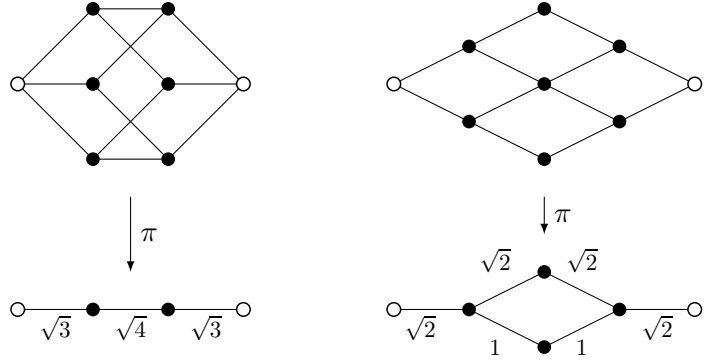

\paragraph{Algebraic characterization}
Our aim in this section is to cast Feder's construction in an algebraic framework. 
Here, we adopt the explicit many-boson quantum walk model used by Gamble \etal \cite{gfzjc10} 
and Smith \cite{smith10}. The Hamiltonian of the $k$-boson quantum walk in this model is given by
\begin{equation} \label{eqn:hamiltonian}
H_{kB} = - \left[\frac{1}{k!} \sum_{\sigma \in S_{k}} P_{\alpha}\right] A(G^{\ecart k}),
\end{equation}
where $S_{k}$ is the symmetric group of all permutations on $k$ elements.
%\mnote{$H_{kB}$ is Hermitian}
Each permutation $\sigma \in S_{k}$ induces the following natural group action on the elements of $V^{k}$,
\begin{equation}
\sigma \circ (x_{1},x_{2},\ldots,x_{k}) = (x_{\sigma(1)},x_{\sigma(2)},\ldots,x_{\sigma(k)}).
\end{equation}
We denote the latter simply as $\sigma(x)$, whenever $x = (x_{1},\ldots,x_{k})$.
So, the permutation matrix $P_{\sigma}$ is an $|V|^{k} \times |V|^{k}$ matrix defined as:
\begin{equation} \label{eqn:p-sigma}
\bra{x}P_{\sigma}\ket{y} = \iverson{y = \sigma(x)}.
\end{equation}

The time evolution of the $k$-boson quantum walk is then given by $U_{kB} = e^{-itH_{kB}}$.
This description captures the intuition that each boson is performing a quantum walk on its own copy 
of the graph $G$ but collectively they are performing a quantum walk on $G^{\ecart k}$.
Next, we show that the {\em symmetrization} operator in Equation (\ref{eqn:hamiltonian})
induces an equitable partition on $G^{\ecart k}$.

\begin{lemma} \label{lemma:symm-equitable}
Let $G=(V,E)$ be a graph and $k \ge 1$ be an integer. Then, the operator
\begin{equation} \label{eqn:symmetrizer}
\SSS = \frac{1}{k!}\sum_{\sigma \in S_{k}} P_{\sigma},
\end{equation}
which acts on the set $V^{k}$, defines an equitable partition $\pi$ of $G^{\ecart k}$.
Moreover, $\mathbb{S}$ equals $QQ^{T}$, 
where $Q$ is the normalized partition matrix of $\pi$.
\end{lemma}

\par\noindent{\em Remark}:
Osborne \cite{o06} considered a related operator which includes an alternating permutation sign 
in the summation. Both operators correspond to symmetrization or skew-symmetrization in a 
symmetric or exterior vector spaces, respectively (see \cite{dummit-foote}, page 452).\\ 

\prf
Consider a vertex partition $\pi = \biguplus_{x} \OO_{x}$ of the product graph $G^{\ecart k}$ defined by the cells
\begin{equation}
\OO_{x} = \{y \in V^k : \exists \sigma \in S_{k},\sigma(x) = y\}.
\end{equation}
Each cell $\OO_{x}$ is an orbit of $S_{k}$ acting on the vertex set $V(G^{\ecart k}) = V^{k}$.
To show $\pi$ is equitable, let $x$ and $y$ be adjacent vertices in $G^{\ecart k}$. 
This implies that there is a unique index $i$ for which $x_{i}$ is adjacent to $y_{i}$ in $G$ 
and $x_{j} = y_{j}$ for all other $j \neq i$ (this can be shown using induction on $k$).
Now, let $S$ be the collection of indices where $x_{i}$ appears in $x$; note $i \in S$. 
Consider a permutation $\tau$ which swaps $i$ with $j \in S \setminus \{i\}$.
Then, $x$ is also adjacent to $\tau(y)$; moreover, $\tau(y) \in \OO_{y}$. 
Thus, $x$ has $|S|$ neighbors in $\OO_{y}$. 
Since $x$ is an arbitrary element of $\OO_{x}$, every element in $\OO_{x}$ has $|S|$ neighbors in $\OO_{y}$.
This shows $\pi$ is equitable.

From the definition of $P_{\sigma}$ in Equation (\ref{eqn:p-sigma}), we have:
\begin{equation}
\bra{x}\left[ \frac{1}{k!}\sum_{\sigma} P_{\sigma}\right]\ket{y}
=
\frac{1}{k!} \sum_{\sigma} \iverson{y = \sigma(x)}
=
\frac{1}{k!} |\Stab(x)| \iverson{y \in \OO_{x}}
= 
\frac{1}{|\OO_{x}|} \iverson{y \in \OO_{x}}
\end{equation}
where $\Stab(x) = \{\sigma \in S_{k}: \sigma(x) = x\}$ is the stabilizer of $x$, which is the set
of permutations which fix $x$.
The last equality follows from $|\OO_{x}||\Stab(x)| = k!$, since the size of the orbit of $x$
is the index of the stabilizer subgroup of $x$ (see Hungerford \cite{hungerford}, Theorem 4.3, page 89).
On the other hand, the partition matrix $Q$ of $\pi$ is defined as 
$\bra{x}Q\ket{j} = |V_{j}|^{-1/2}\iverson{x \in V_{j}}$.
Thus, we have
\begin{equation}
\bra{x}QQ^{T}\ket{y} 
=
\frac{1}{|\OO_{x}|} \iverson{y \in \OO_{x}}.
\end{equation}
This proves our second claim that $\SSS = QQ^{T}$.
\qed\\

%%%%%%%%%%%%%%%%%%%%%%%%%%%%%%%%%%%%%%%%%%%%%%%%%%%%%%%%%%%%%%
%% Feder construction via many-particle quantum walk
%% P3(2-boson) -> P3(Box 2) -> diamond D6
%%%%%%%%%%%%%%%%%%%%%%%%%%%%%%%%%%%%%%%%%%%%%%%%%%%%%%%%%%%%%%
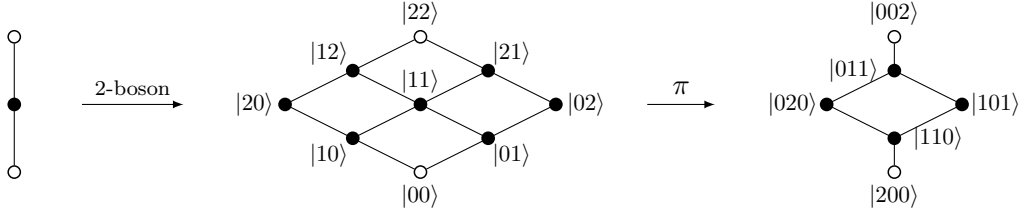
\begin{figure}[t]
\begin{center}
\begin{tikzpicture}[scale=0.9]

% p3 with 2 bosons
\draw (-6,0)--(-6,1)--(-6,2);
% pst nodes
\foreach \y in {0,2}
{
	\node at (-6, \y)[circle, fill=white][scale=0.5]{};
	\draw[line width=0.2mm] (-6, \y) circle (0.09cm);
}
\node at (-6, 1)[circle, fill=black][scale=0.5]{};

% squiggly arrow
\draw[>=latex,->] (-5,1)--(-3.5,1);
\node at (-4.25,1.25){\scriptsize $2$-boson};

% p3 box p3 (vertical orientation)
\draw (0,0)--(-1,0.5)--(-2,1)--(-1,1.5)--(0,2)--(1,1.5)--(2,1)--(1,0.5)--(0,0);
\draw (-1,0.5)--(0,1)--(1,1.5);
\draw (1,0.5)--(0,1)--(-1,1.5);

\foreach \x in {-2, 0, 2}
	\node at (\x, 1)[circle, fill=black][scale=0.5]{};
\foreach \x in {-1, 1}
	\foreach \y in {0.5, 1.5}
		\node at (\x, \y)[circle, fill=black][scale=0.5]{};
% pst nodes
\foreach \y in {0,2}
{
	\node at (0, \y)[circle, fill=white][scale=0.5]{};
	\draw[line width=0.2mm] (0, \y) circle (0.09cm);
}
%\node at (-2, 1)[circle, fill=gray][scale=0.5]{};
%\node at (2, 1)[circle, fill=gray][scale=0.5]{};

% vertex labels
\node at (0,-0.35)[scale=0.8]{$\ket{00}$};
\node at (0,2.35)[scale=0.8]{$\ket{22}$};
\node at (-1.35,0.25)[scale=0.8]{$\ket{10}$};
\node at (1.35,0.25)[scale=0.8]{$\ket{01}$};
\node at (-1.35,1.75)[scale=0.8]{$\ket{12}$};
\node at (1.35,1.75)[scale=0.8]{$\ket{21}$};
\node at (-2.45,1)[scale=0.8]{$\ket{20}$};
\node at (2.45,1)[scale=0.8]{$\ket{02}$};
\node at (0,1.35)[scale=0.8]{$\ket{11}$};

% diamond graph d6
\draw (7,0)--(7,0.5)--(6,1)--(7,1.5)--(7,2);
\draw (7,0.5)--(8,1)--(7,1.5);

\foreach \y in {0.5, 1.5}
	\node at (7, \y)[circle, fill=black][scale=0.5]{};
\foreach \x in {6, 8}
	\node at (\x, 1)[circle, fill=black][scale=0.5]{};
% pst nodes
\foreach \y in {0, 2}
{
	\node at (7, \y)[circle, fill=white][scale=0.5]{};
	\draw[line width=0.2mm] (7, \y) circle (0.09cm);
}
%\node at (8, 1)[circle, fill=gray][scale=0.5]{};

% vertex labels
\node at (7,-0.35)[scale=0.8]{$\ket{200}$};
\node at (7,2.35)[scale=0.8]{$\ket{002}$};
\node at (5.5,1)[scale=0.8]{$\ket{020}$};
\node at (8.5,1)[scale=0.8]{$\ket{101}$};
\node at (7.65,0.5)[scale=0.8]{$\ket{110}$};
\node at (6.4,1.5)[scale=0.8]{$\ket{011}$};

% quotient arrow
\draw[>=latex,->] (3.35,1)--(4.35,1);
\node at (3.85,1.25){$\pi$};

\end{tikzpicture}
\vspace{.1in}
\caption{The $2$-boson walk on $P_{3}$, its Cartesian product representation $P_{3}^{\ecart 2}$, 
and the Feder {\em diamond} graph $D_{6} = \FF{P_{3}}{2} \cong P_{3}^{\ecart 2}/\pi$. 
Antipodal PST occur throughout between vertices marked white.
}
\vspace{.1in}
\label{fig:diamond}
\end{center}
\hrule
\end{figure}

\par\noindent
By Lemma \ref{lemma:symm-equitable}, the unitary evolution $U_{kB}$ of the $k$-boson quantum walk 
admits a simpler description.

\begin{lemma} (see \cite{smith10})
The unitary evolution of the $k$-boson quantum walk on a graph $G$ using the Hamiltonian
$H_{kB} = - \left[\frac{1}{k!} \sum_{\sigma \in S_{k}} P_{\alpha}\right] A(G^{\ecart k})$,
is given by
\begin{equation}
U_{kB} = \left[\frac{1}{k!}\sum_{\sigma} P_{\sigma}\right] (e^{it A(G)})^{\otimes k}.
\end{equation}
\end{lemma}
\prf
First, we note that $(\frac{1}{d} J_{d})^{m} = \frac{1}{d} J_{d}$, for any $d,m \ge 1$.
Let $\SSS = \frac{1}{k!}\sum_{\sigma} P_{\sigma}$ be the ``symmetrizing'' operator defined 
in Equation (\ref{eqn:symmetrizer}).
By Lemma \ref{lemma:equitable-partition} and Lemma \ref{lemma:symm-equitable}, we have $\SSS = QQ^{T}$ 
and it is a block diagonal matrix containing all-one submatrices. 
The block diagonal property of $\SSS$ implies that $\SSS^{m} = \SSS$, for any $m \ge 1$. 
Moreover, again by Lemma \ref{lemma:equitable-partition}, $\SSS$ commutes with $A(G^{\ecart k})$.
Therefore, 
\begin{eqnarray}
U_{kB} 	& = & \exp(-itH_{kB}) = \exp(it \ \SSS A(G^{\ecart k}))  \\
	& = & \sum_{m=0}^{\infty} \frac{(it)^m}{m!}
		\SSS^{m} A(G^{\ecart k})^{m}, \ \ \
		\mbox{ since $\SSS$ commutes with $A(G^{\ecart k})$ } \\
	& = & \SSS \ \exp(it A(G^{\ecart k})), \ \ \
		\mbox{ since $\SSS^{m} = \SSS$, for $m \ge 1$ }
	%& = & \SSS [\exp(it A(G))]^{\otimes k}.
\end{eqnarray}
This proves the claim since $\exp(it A(G^{\ecart k})) = (e^{it A(G)})^{\otimes k}$.
\qed\\

\par\noindent
The next theorem describes our main algebraic characterization of Feder's construction.
We show that the graph $\FF{G}{k}$ {\em is} a quotient graph of the $k$-fold Cartesian product 
$G^{\ecart k}$. Moreover, it shows if $G$ has perfect state transfer, then so does $\FF{G}{k}$, 
which follows immediately from Theorem \ref{thm:equal-fidelity}.

\begin{theorem} \label{thm:feder}
Let $G=(V,E)$ be a graph and $k$ be a positive integer. Then,
\begin{equation}
\FF{G}{k} \cong G^{\ecart k}/\pi,
\end{equation}
where $\pi$ is an equitable partition of $G^{\ecart k}$ defined by the cells
$\OO_{x} = \{y: \exists \sigma \in S_{k}, \ \sigma(x) = y\}$.
Moreover, if $G$ has perfect state transfer then so does $\FF{G}{k}$,
for any positive integer $k$.
\end{theorem}
\prf
Let $\GG = (\mathcal{V},\mathcal{E})$ be the graph $\FF{G}{k}$ described in 
Definition \ref{def:feder-graph}, where $\mathcal{V}$ is the set of $|V|$-dimensional 
vectors whose entries are non-negative integers that sum to $k$.
For $x \in V^{k}$, let $n[x]$ be a $|V|$-dimensional vector whose $u$-th entry, for $u \in V$,
is given by 
\begin{equation}
n[x]_{u} = |\{i \in [k] : x_{i} = u\}|,
\end{equation}
which is the number of occurrences of vertex $u$ in $x$.
Consider the map $\phi: V(G^{\ecart k}) \rightarrow \mathcal{V}$ defined by $\phi(x) = n[x]$.
By definition of $\OO_{x}$, we have $n[y] = n[x]$ for all $y \in \OO_{x}$.
So, we may view $\phi$ as a mapping from $V(G^{\ecart k}/\pi)$ to $\mathcal{V}$.

Next, we show that $\phi$ is a graph isomorphism between the quotient graph $G^{\ecart k}/\pi$ 
and Feder's graph $\GG$. 
Consider two vertices $\OO_{x}$ and $\OO_{y}$ of the quotient graph $G^{\ecart k}/\pi$ 
whose edge weight between them is $\sqrt{d_{x,y}d_{y,x}}$. 
Here, 
$d_{x,y}$ is the number of neighbors in $\OO_{y}$ that each vertex in $\OO_{x}$ has
and 
$d_{y,x}$ is the number of neighbors in $\OO_{x}$ that each vertex in $\OO_{y}$ has.

Let $\phi(\OO_{x}) = n[x]$ and $\phi(\OO_{y}) = n[y]$.
If $x$ and $y$ are adjacent in the product graph $G^{\ecart k}$, 
then $x$ and $y$ differ in exactly one coordinate $i$, where $x_{i}$ and $y_{i}$ are adjacent in $G$, 
and agree in the other coordinates.
Suppose $x_{i}=u$ and $y_{i}=v$ with $u \neq v$ but $u$ is adjacent to $v$ in $G$.
Then, $n[y]_{u} = n[x]_{u} - 1$ and $n[y]_{v} = n[x]_{v} + 1$.
By Definition \ref{def:feder-graph}, the edge weight between $n[x]$ and $n[y]$ in $\GG$ is given by
\begin{equation}
\omega(n[x],n[y]) = \sqrt{n[x]_{u}(n[x]_{v} + 1)} 
\end{equation}
which equals to
\begin{equation}
\omega(\OO_{x},\OO_{y}) = \sqrt{d_{x,y}d_{y,x}}, 
\end{equation}
since $d_{x,y} = n[x]_{u}$ (the number of ways to replace $u$ with $v$)
and $d_{y,x} = n[y]_{v}+1$ (the number of ways to replace $v$ with $u$).
This shows that $\FF{G}{k} \cong G^{\ecart k}/\pi$.
\qed\\

\par\noindent
The next theorem shows a composition theorem for Feder's operator $\FF{G}{k}$.
We will use this to describe a {\em reduction} method from one perfect state transfer graph to another
by combining and alternating lifting and quotient operations.

\begin{theorem} \label{thm:composition}
For a given graph $G$ and integers $m_{1},m_{2} \ge 1$, 
let $\pi_{1}$ be an equitable partition of $G^{\ecart m_{1}}$ and
let $\pi_{2}$ be an equitable partition of $(G^{\ecart m_{1}}/\pi_{1})^{\ecart m_{2}}$. 
Then, there is an equitable partition $\pi_{3}$ of $G^{\ecart (m_{1}m_{2})}$ where
\begin{equation}
(G^{\ecart m_{1}}/\pi_{1})^{\ecart m_{2}}/\pi_{2} \cong G^{\ecart (m_{1}m_{2})}/\pi_{3}.
\end{equation}
\end{theorem}
\prf
Let $Q_{1}$ and $Q_{2}$ be the (normalized) partition matrices corresponding to $\pi_{1}$ and $\pi_{2}$,
respectively.
The adjacency matrix of $(G^{\ecart m_{1}}/\pi_{1})^{\ecart m_{2}}$ is given by
\begin{equation}
\sum_{k=1}^{m_{2}} (I \otimes \ldots \otimes 
	\overbrace{Q_{1}^{T}A(G^{\ecart m_{1}})Q_{1}}^{\mbox{\scriptsize $k$-th position}} 
	\otimes \ldots I),
\end{equation}
since $A(G^{\ecart m_{1}}/\pi_{1}) = Q_{1}^{T} A(G^{\ecart m_{1}}) Q_{1}$. 
By expressing the identity matrices as $Q_{1}^{T}Q_{1}$ and factoring it out from both sides, we get
\begin{equation}
(Q_{1}^{T})^{\otimes m_{2}}
\left[
\sum_{k=1}^{m_{2}} (I \otimes \ldots \otimes 
	\overbrace{A(G^{\ecart m_{1}})}^{\mbox{\scriptsize $k$-th position}} 
	\otimes \ldots I)
\right]
Q_{1}^{\otimes m_{2}}.
\end{equation}
The last equation yields 
\begin{equation}
(Q_{1}^{\otimes m_{2}})^{T} \left[A(G^{\ecart m_{1}})^{\ecart m_{2}}\right] Q_{1}^{\otimes m_{2}}
=
(Q_{1}^{\otimes m_{2}})^{T} A(G^{\ecart (m_{1}m_{2})}) Q_{1}^{\otimes m_{2}}.
\end{equation}
Thus, the adjacency matrix of 
$(G^{\ecart m_{1}}/\pi_{1})^{\ecart m_{2}}/\pi_{2}$
is given by
\begin{equation}
Q_{2}^{T} (Q_{1}^{\otimes m_{2}})^{T} A(G^{\ecart (m_{1}m_{2})}) Q_{1}^{\otimes m_{2}} Q_{2},
\end{equation}
which proves the claim and shows $\pi_{3}$ is defined by the partition matrix
$Q_{1}^{\otimes m_{2}}Q_{2}$.
\qed\\

\par\noindent{\em Remark}:
Using Theorem \ref{thm:composition}, the perfect state transfer graphs described in \cite{f06} 
arguably are all quotients of the $n$-cube derived using different equitable partitions.
For example, the graph shown in Figure \ref{fig:diamond} is derived from the $4$-cube since
$P_{3}^{\ecart 2}/\pi_{1} \cong (K_{2}^{\ecart 2}/\pi_{2})^{\ecart 2}/\pi_{1} \cong K_{2}^{\ecart 4}/\pi_{3}$.

%%%%%%%%%%%%%%%%%%%%%%%%%%%%%%%%%%%%%%%%%%%%%%%%%%%%%%%%%%%%%%%%%%%%%%%%%%%%%%%%%%%%%%%%%%%%%%%%%%%%%%%%%%%%%%%%%%

\section{Generalizations}

%%%%%%%%%%%%%%%%%%%%%%%%%%%%%%%%%%%%%%%%%%%%%%%%%%%%%%%%%%%%%%
%% product of PST and periodic graph: K2 Box Q3(+011)
%%%%%%%%%%%%%%%%%%%%%%%%%%%%%%%%%%%%%%%%%%%%%%%%%%%%%%%%%%%%%%
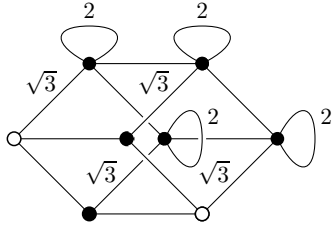
\begin{figure}[t]
\begin{center}
\begin{tikzpicture}

\draw (6,0)--(7.5,0);
\draw (7,1)--(8.5,1);
\draw (7,-1)--(8.5,-1);

% two collapsed 3-cube BGS
\foreach \x in {0,1.5}	% shift x by a certain amount
\foreach \y in {0}	% could shift y also be a certain amount if necessary
{
\draw (6+\x,0+\y)--(7+\x,1+\y)--(8+\x,0+\y)--(7+\x,-1+\y)--(6+\x,0+\y);

\node at (8+\x, 0+\y)[circle, fill=black][scale=0.5]{};
\node at (7+\x, 1+\y)[circle, fill=black][scale=0.5]{};

% pst nodes
\node at (6+\x, 0+\y)[circle, fill=white][scale=0.5]{};
\draw[line width=0.2mm] (6+\x, 0+\y) circle (0.09cm);
\node at (7+\x, -1+\y)[circle, fill=white][scale=0.5]{};
\draw[line width=0.2mm] (7+\x, -1+\y) circle (0.09cm);

\draw (7+\x,1+\y) .. controls (6.5+\x,1.25+\y) and (6.5+\x,1.5+\y) .. (7+\x,1.5+\y) .. controls (7.5+\x,1.5+\y) and (7.5+\x,1.25+\y) .. (7+\x,1+\y);
\draw (8+\x,0+\y) .. controls (8.25+\x,0.5+\y) and (8.5+\x,0.5+\y) .. (8.5+\x,0+\y) .. controls (8.5+\x,-0.5+\y) and (8.25+\x,-0.5+\y) .. (8+\x,0+\y);

% weights on self-loops
\node at (7+\x, 1.7+\y)[scale=0.8]{$2$};
\node at (8.65+\x, 0.3+\y)[scale=0.8]{$2$};

% other non-unit weights 
\node at (6.35+\x, 0.75+\y)[scale=0.8]{$\sqrt{3}$};
\node at (7.15+\x, -0.45+\y)[scale=0.8]{$\sqrt{3}$};
}

\draw (8,0)--(9.5,0);

% self-loop to foreground
\tikzstyle myBG=[line width=3pt,opacity=1.0]
\foreach \x in {0}
\foreach \y in {0}
{
	% wipeout
	\draw[white,myBG] (8+\x,0+\y) .. controls (8.25+\x,0.5+\y) and (8.5+\x,0.5+\y) .. (8.5+\x,0+\y) .. controls (8.5+\x,-0.5+\y) and (8.25+\x,-0.5+\y) .. (8+\x,0+\y);
	% redraw
	\draw (8+\x,0+\y) .. controls (8.25+\x,0.5+\y) and (8.5+\x,0.5+\y) .. (8.5+\x,0+\y) .. controls (8.5+\x,-0.5+\y) and (8.25+\x,-0.5+\y) .. (8+\x,0+\y);
	\node at (8+\x, 0+\y)[circle, fill=black][scale=0.5]{};
}

\node at (7.5,0)[circle, fill=black][scale=0.5]{};
\node at (7,-1)[circle, fill=black][scale=0.5]{};

% two diagonal edges to foreground: wipeout & redraw
\draw[white, myBG] (7.5, 0)--(8.5, 1);
\draw[white, myBG] (7.5, 0)--(8.5,-1);
\draw (7.5, 0)--(8.5, 1);
\draw (7.5, 0)--(8.5,-1);
\node at (7.5, 0)[circle, fill=black][scale=0.5]{};
\node at (8.5, 1)[circle, fill=black][scale=0.5]{};
% pst node redraw
\node at (8.5,-1)[circle, fill=white][scale=0.5]{};
\draw[line width=0.2mm] (8.5,-1) circle (0.09cm);

\end{tikzpicture}
\vspace{.1in}
\caption{The perfect state transfer graph $K_{2} \cart (X(\ZZ_{2}^{3},\{001,010,100,011\})/\pi)$,
where the latter is a Cartesian product of $K_{2}$ with a periodic graph.
}
\vspace{.1in}
\label{fig:mix-product}
\end{center}
\hrule
\end{figure}

\subsection{Inhomogeneous products}

Note that Feder's construction is based on taking the Cartesian product of a single perfect state
transfer graph with itself followied by a quotient operation.
Here, we extend this construction by using distinct perfect state transfer and periodic graphs
in the product and by allowing the quotient operations to {\em alternate} with the product.
But first, we show a composition theorem for this more general construction (similar to Theorem \ref{thm:composition}).

\begin{theorem} \label{thm:mixed-product}
For $n \in \NN$ and for each $k \in [n]$, let $G_{k}$ be a graph and $\pi_{k}$ be an associated equitable partition. 
Then, there is an equitable partition $\pi$ so that
\begin{equation}
\Box_{k=1}^{n} (G_{k}/\pi_{k}) = (\Box_{k=1}^{n} G_{k})/\pi.
\end{equation}
Moreover, if $Q_{k}$ is the partition matrix of $\pi_{k}$, then
$\bigotimes_{k=1}^{n} Q_{k}$ is the partition matrix of $\pi$.
\end{theorem}
\prf
Let $Q_{k}$ be the normalized partition matrix of $\pi_{k}$.
The adjacency matrix of $G_{k}/\pi_{k}$ is defined by $Q_{k}^{T}A(G_{k})Q_{k}$.
Thus, the adjacency matrix of $\Box_{k} (G_{k}/\pi_{k})$ is 
\begin{equation}
\sum_{k=1}^{n} (I \otimes \ldots \otimes 
	\overbrace{Q_{k}^{T}A(G_{k})Q_{k}}^{\mbox{\scriptsize $k$-th position}} 
	\otimes \ldots \otimes I).
\end{equation}
Now, replace each $I$ in the term above by $Q_{j}^{T}Q_{j}$ if it is in position $j \neq k$.
This gives us
\begin{equation}
\sum_{k=1}^{n} (Q_{1}^{T}Q_{1} \otimes \ldots \otimes 
	\overbrace{Q_{k}^{T}A(G_{k})Q_{k}}^{\mbox{\scriptsize $k$-th position}} 
	\otimes \ldots \otimes Q_{n}^{T}Q_{n}).
\end{equation}
Factoring the common terms $Q_{k}^{T}$ on the left and $Q_{k}$ on the right, we get
\begin{equation}
\left(\bigotimes_{k=1}^{n} Q_{k}^{T}\right) 
	\sum_{k=1}^{n} (I \otimes \ldots \otimes 
		\overbrace{A(G_{k})}^{\mbox{\scriptsize $k$-th position}} 
		\otimes \ldots \otimes I)
	\left(\bigotimes_{k=1}^{n} Q_{k}\right).
\end{equation}
This yields
\begin{equation}
\left(\bigotimes_{k=1}^{n} Q_{k}\right)^{T} A(\Box_{k=1}^{n} G_{k}) \left(\bigotimes_{k=1}^{n} Q_{k}\right).
\end{equation}
which shows that $Q = \bigotimes_{k=1}^{n} Q_{k}$ is the partition matrix of $\pi$.
\qed\\

\noindent
The following corollary extends Feder's operator $\FF{G}{k}$ which is based on a single graph $G$.
Here, we take a product of different graphs $G_{k}$ (and their quotients $G_{k}/\pi_{k}$) and allow
both perfect state transfer and periodic graphs with commensurable times.

\begin{corollary} \label{cor:generalized-feder}
Let $n \ge 1$ be an integer. For $k \in [n]$, 
let $G_{k}$ be a graph with perfect state transfer between vertices $a_{k}$ and $b_{k}$ at time $t$
($G_{k}$ is periodic, if $a_{k} = b_{k}$), where $a_{k} \neq b_{k}$ for at least one $k$.
Let $\pi_{k}$ be an equitable distance partition of $G_{k}$ with respect to $a_{k}$ and $b_{k}$.
Then 
\begin{equation}
\Box_{k=1}^{n} (G_{k}/\pi_{k}) \cong (\Box_{k=1}^{n} G_{k})/\pi
\end{equation}
has perfect state transfer between $(a_{1},\ldots,a_{n})$ and $(b_{1},\ldots,b_{n})$ at time $t$.
Here, $\pi$ is an equitable partition of~ $\Box_{k} G_{k}$ defined by the partition matrix $\otimes_{k} Q_{k}$,
where $Q_{k}$ is the partition matrix of $\pi_{k}$.
\end{corollary}

We show an example of how to build new perfect state transfer graphs using Corollary \ref{cor:generalized-feder}. 
For this, we use the following powerful results on cube-like graphs proved by Bernasconi \etal \cite{bgs08} and 
by Cheung and Godsil \cite{cg11}.

\begin{theorem} (Bernasconi \etal \cite{bgs08} and Cheung-Godsil \cite{cg11}) \label{thm:bgs-cg} \\
Let $G = X(\ZZ_{2}^n,S)$ be the Cayley graph on $\ZZ_{2}^{n}$ with generating set $S$
and let $\omega_{S} = \sum_{a \in S} a$ be the sum of the elements in $S$. 
Let $M$ be the $n \times |S|$ matrix with elements of $S$ as columns and whose row space is called
the {\em code} of $G$. Also, let $D$ be the greatest common divisor of the weights of the codewords of $G$.
Then, the following holds:
\begin{enumerate}
\item If $\omega_{S} \neq 0$, then $G$ has perfect state transfer from $0$ to $\omega_{S}$ at time $t=\pi/2$.
\item If $\omega_{S}=0$, then $G$ has perfect state transfer at time $t=\pi/4$ if and only if $D=2$ and
	the code of $G$ is self-orthogonal.
\end{enumerate}
\end{theorem}

\par\noindent{\em Remark}:
Let $G_{k}=X(\ZZ_{2}^{n},S_{k})$ be any collection of cube-like graphs defined in Theorem \ref{thm:bgs-cg}, 
where at least one satisfies $\sum_{a \in S_{k}} a \neq 0$. This guarantees that at least one graphs has
``antipodal'' perfect state transfer at time $\pi/2$, while the others might be periodic at time $\pi/2$.
By Corollary \ref{cor:generalized-feder}, we know $\Box_{k} (G_{k}/\pi_{k}) \cong (\Box_{k} G_{k})/\pi$ 
has perfect state transfer, for any collection of equitable partitions $\{\pi_{k}\}$.
A simple example of this construction is given in Figure \ref{fig:mix-product}.

%%%%%%%%%%%%%%%%%%%%%%%%%%%%%%%%%%%%%%%%%%%%%%%%%%%%%%%%%%%%%%%%%%%%%%%%%%%%%%%%%%%%%%%%%%%%%%%%%%%%%%%%%%%%%%%%%%

\subsection{Reductions} \label{section:reductions}

%%%%%%%%%%%%%%%%%%%%%%%%%%%%%%%%%%%%%%%%%%%%%%%%%%%%%%%%%%%%%%
%% the lift-and-quotient technique
%% source: Michael Landry, July 2011
%%%%%%%%%%%%%%%%%%%%%%%%%%%%%%%%%%%%%%%%%%%%%%%%%%%%%%%%%%%%%%
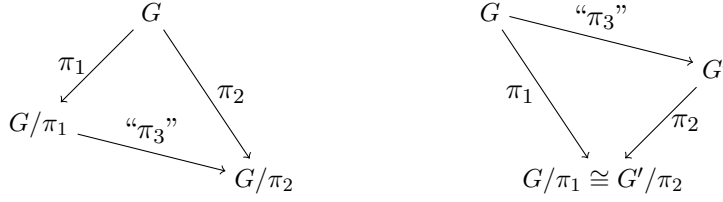
\begin{figure}[t]
%\hrule
%\vspace{.15in}
\begin{center}
\begin{tikzpicture}[scale=0.75]
% lift-and-quotient
\node(P1) at (0,0)[scale=0.9]{$G/\pi_{1}$};
\node(P0) at (2,2)[scale=0.9]{$G$};
\node(P2) at (4,-1)[scale=0.9]{$G/\pi_{2}$};

\draw
(P0) edge[->] node[left]{$\pi_{1}$} (P1)
(P0) edge[->] node[right]{$\pi_{2}$} (P2)
(P1) edge[->] node[above]{\mbox{``$\pi_{3}$''}} (P2);

% quotient-and-lift
\node(Q0) at (8,2)[scale=0.9]{$G$};
\node(Q1) at (10,-1)[scale=0.9]{$G/\pi_{1} \cong G'/\pi_{2}$};
\node(Q2) at (12,1)[scale=0.9]{$G'$};

\draw
(Q0) edge[->] node[left]{$\pi_{1}$} (Q1)
(Q2) edge[->] node[right]{$\pi_{2}$} (Q1)
(Q0) edge[->] node[above]{\mbox{``$\pi_{3}$''}} (Q2);

\end{tikzpicture}
\vspace{.1in}
\caption{
(i) Lift-and-quotient:
if $G/\pi_{1}$ has PST, then $G$ has PST; which implies $G/\pi_{2}$ has PST. 
So, $G/\pi_{1}$ reduces to $G/\pi_{2}$ via $\mbox{``$\pi_{3}$''}$.
(ii) Quotient-and-lift:
if $G$ has PST, then $G/\pi_{1}$ has PST; which implies $G'$ has PST if $G'/\pi_{2} \cong G/\pi_{1}$.
So, $G$ reduces to $G'$ via $\mbox{``$\pi_{3}$''}$.
}
\vspace{.1in}
\label{fig:reductions}
\end{center}
\hrule
\end{figure}

In this section, we describe reductions between perfect state transfer graphs obtained
from alternating a {\em lifting} move (from a quotient graph $G/\pi$ to a graph $G$,
for an equitable partition $\pi$) and a {\em quotient} move (from the graph $G$ to its quotient
graph $G/\pi$, for a possibly different equitable partition). 
By interchanging the order of these two operations, we get a quotient-and-lift reduction
or a lift-and-quotient reductions.
We illustrate these two types of reductions in Figure \ref{fig:reductions}.

\begin{figure}[t]
\begin{center}
\begin{tikzpicture}
%% diamond d6
\foreach \s in {-1}
{
\draw (0+\s,0)--(1+\s,0)--(2+\s,0.5)--(3+\s,0)--(4+\s,0);
\draw (1+\s,0)--(2+\s,-0.5)--(3+\s,0);

% pst nodes
\foreach \x in {0,4}
{
	\node at (\x+\s, 0)[circle, fill=white][scale=0.5]{};
	\draw[line width=0.2mm] (\x+\s, 0) circle (0.09cm);
}

\foreach \x in {1,3}
	\node at (\x+\s, 0)[circle, fill=black][scale=0.5]{};
\foreach \y in {-0.5,0.5}
	\node at (2+\s, \y)[circle, fill=black][scale=0.5]{};

\node at (0.5+\s, 0.25)[scale=0.8]{$\sqrt{2}$};
\node at (3.5+\s, 0.25)[scale=0.8]{$\sqrt{2}$};
\node at (1.3+\s, 0.5)[scale=0.8]{$\sqrt{2}$};
\node at (2.65+\s, 0.5)[scale=0.8]{$\sqrt{2}$};
\node at (1.35+\s, -0.5)[scale=0.8]{$1$};
\node at (2.65+\s, -0.5)[scale=0.8]{$1$};
}

%% lifted diamond
\draw (4,2)--(5,2)--(6,2.6)--(7,2)--(8,2);
\draw (5,2)--(6,1.5)--(7,2);
\draw (5,2)--(6,2.3)--(7,2);

% pst nodes
\foreach \x in {4,8}
{
	\node at (\x, 2)[circle, fill=white][scale=0.5]{};
	\draw[line width=0.2mm] (\x, 2) circle (0.09cm);
}

\foreach \x in {5,7}
	\node at (\x, 2)[circle, fill=black][scale=0.5]{};
\foreach \y in {1.5, 2.3, 2.6}
	\node at (6, \y)[circle, fill=black][scale=0.5]{};

\node at (4.5, 2.25)[scale=0.8]{$\sqrt{2}$};
\node at (7.5, 2.25)[scale=0.8]{$\sqrt{2}$};

%% weighted P5 with PST
\draw (6,-1)--(10,-1);

% pst nodes
\foreach \x in {6, 10}
{
	\node at (\x, -1)[circle, fill=white][scale=0.5]{};
	\draw[line width=0.2mm] (\x, -1) circle (0.09cm);
}

\foreach \x in {7,8,9}
	\node at (\x, -1)[circle, fill=black][scale=0.5]{};
\node at (6.5, -0.75)[scale=0.8]{$\sqrt{2}$};
\node at (7.5, -0.75)[scale=0.8]{$\sqrt{3}$};
\node at (8.5, -0.75)[scale=0.8]{$\sqrt{3}$};
\node at (9.5, -0.75)[scale=0.8]{$\sqrt{2}$};

% the quotient arrows

\node(X1) at (3.5,-0.25){};
\node(X2) at (5.85,-0.85){};
\draw
(X1) edge[->] node[above]{\mbox{``$\pi_{3}$''}} (X2);

\node(Y1) at (6.5,1.5){};
\node(Y2) at (8,-0.5){};
\draw
(Y1) edge[->] node[right]{$\pi_{2}$} (Y2);

\node(Z1) at (5.5,1.5){};
\node(Z2) at (3.5,0.25){};
\draw
(Z1) edge[->] node[above]{$\pi_{1}$} (Z2);
\end{tikzpicture}
\vspace{.1in}
\caption{The lift-and-quotient reduction:
$D_{6} = P_{3}^{\ecart 2}/\pi$ is {\em lifted} to the top graph $G$ (via the ``inverse'' of $\pi_{1}$) 
whose ``other'' quotient is $D_{6}/\pi_{3}$. We infer $G$ has PST for ``free''.
}
\vspace{.1in}
\label{fig:lift-quotient}
\end{center}
\hrule
\end{figure}
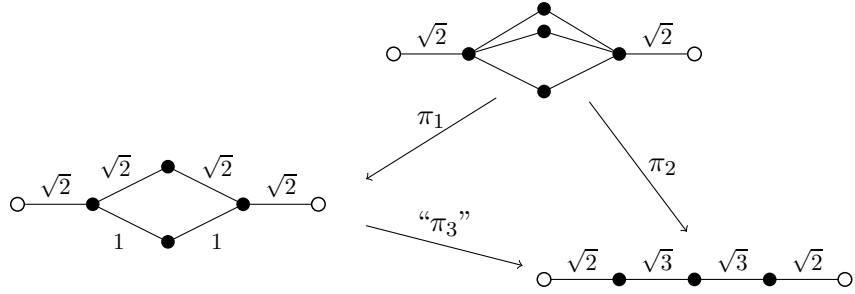

As a simple example, consider the diamond graph $D_{6}$ from Figure \ref{fig:diamond}.
There is a lift-and-quotient reduction from $D_{6}$ to $\PS_{5}$ given by
\begin{equation}
D_{6} \nearrow G \searrow \PS_{5} 
\end{equation}
where $G$ is the graph obtained from attaching two vertices onto $K_{2,3}$ (each with edge weight $\sqrt{2}$).
This reduction is depicted in Figure \ref{fig:lift-quotient}. Note we get PST on $G$ for ``free''.
An alternate lift-and-quotient reduction based on Theorem \ref{thm:composition} is given by
\begin{equation}
D_{6} = P_{3}^{\ecart 2}/\pi_{1} \nearrow (K_{2}^{\ecart 2}/\pi_{2})^{\ecart 2}/\pi_{1} 
	\cong K_{2}^{\ecart 4}/\pi_{3} \searrow \PS_{5} 
\end{equation}
where here $G$ is the $4$-cube $Q_{4} = K_{2}^{\ecart 4}$.

%%%%%%%%%%%%%%%%%%%%%%%%%%%%%%%%%%%%%%%%%%%%%%%%%%%%%%%%%%%%%%%%%%%%%%%%%%%%%%%%%%%%%%%%%%%%%%%%%%%%%%%%%%%%%%%%%%

\paragraph{Irreducible graphs}
Godsil's question in \cite{godsil-survey} is closely related to an observation of Kay \cite{kay10} 
that any weighted path with perfect state transfer must have mirror-symmetric weights.
Given the construction described in Section \ref{section:xeg}, it is natural to ask if
there is a class of graphs for which perfect state transfer implies the automorphism property.
Let $G=(V,E)$ be a graph with perfect state transfer between vertices $a$ and $b$. 
For each vertex $x \in V$, let $d_{a}(x)$ (respectively, $d_{b}(x)$) be the distance of $x$ from
$a$ (respectively, $b$). To each vertex $x$, we assign the distance-pair $d_{a,b}(x)=(d_{a}(x),d_{b}(x))$ 
of $x$ from both $a$ and $b$. 
We say $G$ is {\em distance-minimal} with respect to vertices $a$ and $b$ if each vertex has a unique 
distance-pair, that is, for $x \neq y$, we have $d_{a,b}(x) \neq d_{a,b}(y)$.
Alternatively, we say a graph $G_{1}$ is {\em reducible} to $G_{2}$ (with respect to vertices $a$ and $b$)
if there is a lift-and-quotient or a quotient-and-lift reduction from $G_{1}$ to $G_{2}$ which places vertices 
$a$ and $b$ in singleton cells, so that $G_{2}$ has fewer vertices than $G_{1}$.
We call a graph {\em quotient-minimal} if it is not reducible to any other graph.
Let us call a graph {\em minimal} if it is either distance-minimal or quotient-minimal.
Intuitively, if a graph is minimal, it can only have (if any) an automorphism 
switching $a$ and $b$ since the action of permuting vertices at the same distance 
from $a$ or $b$ have been ruled out.

\begin{conjecture}
Let $G$ be a graph with perfect state transfer between vertices $a$ and $b$.
If $G$ is {\em minimal} with respect to $a$ and $b$, then $G$ has an automorphism 
$\tau \in \Aut(G)$ so that $\tau(a)=b$.
\end{conjecture}

%%%%%%%%%%%%%%%%%%%%%%%%%%%%%%%%%%%%%%%%%%%%%%%%%%%%%%%%%%%%%%
%% the lift-and-quotient technique
%% source: Michael Landry, July 2011
%%%%%%%%%%%%%%%%%%%%%%%%%%%%%%%%%%%%%%%%%%%%%%%%%%%%%%%%%%%%%%

%%%%%%%%%%%%%%%%%%%%%%%%%%%%%%%%%%%%%%%%%%%%%%%%%%%%%%%%%%%%%%%%%%%%%%%%%%%%%%%%%%%%%%%%%%%%%%%%%%%%%%%%%%%%%%%%%%

\section{Conclusions}

In this work, we explored perfect state transfer in quantum walks using equitable partitions. 
Our main focus is on a strong equivalence of perfect state transfer between a graph and its 
quotients. Although {\em weaker} forms of this equivalence had appeared earlier, we gave a simple 
and most direct proof which yields a powerful two-way tool (taking lifts and quotients on graphs) 
to study perfect state transfer.

In {\em lifting}, if a perfect state transfer graph is a quotient of another graph, then 
the parent graph also has perfect state transfer. 
We used this to construct graphs with perfect state transfer between two vertices
but which lack automorphism swapping the vertices; hence, answering a question posed by
Godsil in \cite{godsil-survey}. This question is relevant since, prior to this work, all known 
graphs with perfect state transfer admit the automorphism property.

In a {\em quotient} move, if a graph has perfect state transfer graph, then so does its quotient. 
These quotient graphs are obtained by forming various equitable partitions of the original graph. 
We used this to describe Feder's intriguing construction of PST graphs \cite{f06} based on 
many-boson quantum walks.
By adopting an explicit model of $k$-boson quantum walk in \cite{gfzjc10,smith10}, 
we show that Feder's graphs are quotients of a $k$-fold Cartesian product of PST graphs. 
The resulting graphs have perfect state transfer due to the equivalence theorem.
This is related to works by Audenaart \cite{agrr07}, by Osborne \cite{o06}, 
and by Wie\'{s}niak and Markiewicz \cite{wm09} which used algebraic graph theory to provide
explicit connection between multiple and single excitation subspaces under various coupling 
schemes on graphs. 

It would be interesting to find a property of graphs, for which any graph perfect state transfer 
graph with this property must admit an automorphism swapping the two perfect state transfer vertices. 
We leave this as an open question for future work.

%%%%%%%%%%%%%%%%%%%%%%%%%%%%%%%%%%%%%%%%%%%%%%%%%%%%%%%%%%%%%%%%%%%%%%%%%%%%%%%%%%%%%%%%%%%%%%%%%%%%%%%%%%%%%%%%%%

\section*{Acknowledgments}

The research was supported in part by the National Science Foundation grant DMS-1004531
and also by the National Security Agency grant H98230-11-1-0206.
We thank David Feder for kindly describing his construction in \cite{f06}, 
Chris Godsil for very helpful comments on quantum walks, 
Dani ben-Avraham for discussions on bosons,
and the anonymous reviewers for constructive comments which improved the
presentation of this paper.

%%%%%%%%%%%%%%%%%%%%%%%%%%%%%%%%%%%%%%%%%%%%%%%%%%%%%%%%%%%%%%%%%%%%%%%%%%%%%%%%%%%%%%%%%%%%%%%%%%%%%%%%%%%%%%%%%%

\end{document}